\documentclass[%
 reprint,
superscriptaddress,
 amsmath,amssymb,
 prb,
]{revtex4-2}
\usepackage{physics}
\usepackage{float}
\usepackage{lipsum}
\usepackage{graphicx}
\usepackage{dcolumn}
\usepackage{bm}
\usepackage{hyperref}
\usepackage{subcaption}
\usepackage{tikz}
\usetikzlibrary{arrows,decorations.markings,plotmarks}
\usepackage[compat=1.1.0]{tikz-feynman}
\tikzfeynmanset{every vertex/.style={dot},warn luatex=false}
\usepackage{float}						
\usepackage{xcolor}
\usepackage{gensymb}
\def\k{\mathbf{k}}
\def\q{\mathbf{q}}
\def\G{\mathbf{G}}
\def\r{\mathbf{r}}

\def\exciting{{\usefont{T1}{lmtt}{b}{n}exciting}}

\newcommand{\ie}{i.e., }

\usepackage{tabularx}
\usepackage{array}
\newcolumntype{P}[1]{>{\centering\arraybackslash}p{#1}}
\newcolumntype{Y}{>{\centering\arraybackslash}X}

\begin{document}


\title{Phonon-mediated renormalization of exciton energies and absorption spectra in polar semiconductors}
\author{Maximilian Schebek}
\email{m.schebek@fu-berlin.de}

\affiliation{Fachbereich Physik, Freie  Universit\"at  Berlin, D-14195 Berlin, Germany\\}%
\affiliation{Institut f\"ur Physik, Humboldt-Universit\"at zu Berlin, D-12489 Berlin, Germany\\}
\author{Pasquale Pavone}
\affiliation{Institut f\"ur Physik, Humboldt-Universit\"at zu Berlin, D-12489 Berlin, Germany\\}
\author{Claudia Draxl}
\affiliation{Institut f\"ur Physik, Humboldt-Universit\"at zu Berlin, D-12489 Berlin, Germany\\}
\author{Fabio Caruso}
 \affiliation{Institut f\"ur Theoretische Physik und Astrophysik, Christian-Albrechts-Universit\"at zu Kiel, D-24098 Kiel, Germany}
\affiliation{Institut f\"ur Physik, Humboldt-Universit\"at zu Berlin, D-12489 Berlin, Germany\\}

\begin{abstract}
We investigate the influence of vibrational screening on the excitonic and optical properties of solids based on first-principles electronic-structure calculations. We solve the Bethe-Salpeter equation -- the state-of-the-art description of excitons -- by explicitly accounting for phonon-assisted screening effects in the screened Coulomb interaction. In the examples of the polar semiconductors ZnS, MgO, and GaN, the exciton binding energies at the absorption onset are found to be renormalized by a few tens of meV. Similar effects are also found for higher-lying unbound electron-hole pairs, leading to red-shifts of the absorption peaks by up to 50~meV. Our analysis reveals that vibrational screening is dictated by long-range Fr\"ohlich coupling involving polar longitudinal optical phonons, whereas the remaining vibrational degrees of freedom are negligible. Overall, by elucidating the influence of phonon screening on the excitonic states and absorption spectra of these selected ionic semiconductors, this study contributes to advancing the {\it ab initio} methodology and the fundamental understanding of exciton-phonon coupling in solids. 
\end{abstract}
\maketitle

\section{\label{intro} Introduction}
The properties of excitons -- bound electron-hole pairs in photoexcited semiconductors and insulators -- govern the optical response of solids and determine the suitability of materials for applications in optoelectronic devices~\cite{cardona2005fundamentals,Mueller2018-tn} such as UV photo-detectors~\cite{Li2023-sw, Tang2022-un}, light-emitting diodes (LEDs)~\cite{Udagawa2014,Sundaram2013-ac}, and photovoltaic absorbers~\cite{Grancini2012,Kojima2009,Malinkiewicz2013,Stranks2015-ne}. The development of new theoretical and computational frameworks for their description is an important prerequisite for advancing the fundamental understanding of light-matter interactions and disclosing unexplored opportunities for optimizing device performance. The advancements of electronic structure approaches based on density-functional theory and many-body perturbation theory has enabled the formulation of predictive, transferable, and reproducible  workflows for numerical simulations of excitonic and optical properties starting from first principles~\cite{ks_dft_65,hybertsen1986electron,eiguren_epc_2009, Giustino, DESLIPPE20121269}. 

The Bethe-Salpeter equation (BSE) \cite{strinati1988application,rohlf_louie_2000, rohlf_louie_1998} constitutes the state of the art for the {\it ab initio} description of excitons in materials. The BSE accounts for the attractive Coulomb interaction between electrons and holes, which is screened due to the polarization of electronic and vibrational degrees of freedom. However, most BSE implementations only account for electronic screening effects, even though numerous manifestions of exciton-phonon coupling are known, underpinning a wide spectrum of phenomena. Examples include the finite linewidth of excitonic peaks~\cite{rudin_exc_line_1990, Chan2023-jg}, the temperature dependence of absorption spectra~\cite{marini_finite_t_exc_2008,exciton_perovskites}, the formation of excitonic polarons~\cite{Iadonisi1987, giustino_exc_polarons_24_prl}, phonon-assisted optical absorption~\cite{noffsinger_opt_ph_2012,zacharias_ph_opt_2015,Patrick2014}, exciton-phonon scattering in ultrafast dynamics~\cite{helmrich_2021}, and vibrational screening of the electron-hole interaction~\cite{bechstedt2016many,shree_excph_mose2_2018}. Earlier theoretical works based on model Hamiltonians revealed that vibrational coupling can significantly modify excitonic properties by introducing additional screening effects, activating new scattering channels~\cite{tassone_1999}, and leading to the formation of new types of quasiparticles~\cite{Iadonisi1983,Toyozawa_58,yarkony_76, pollmann_1977,mahanti_eff_el_1970, mahanti_eff_el_1972}. Recent advances in the description of exciton-phonon interaction have focused on extending and generalizing the BSE formalism for tackling these phenomena  partially or fully from first principles~\cite{verzelen_exc_pol_2022, bechstedt_ga203, Schleife2018, Park2022, yang_exc_struc_relax_22, chen_exc_ph_20, cudazzo_2020, antonius_exc_ph, paleari_exc_ph_22, Chan2023_exc_life_line, filip_perovksites, alvertis_bse_exc_2023, Gillet2017, Olovsson2019}.

 A particularly striking manifestation of exciton-phonon coupling is the reduction of exciton binding energies caused by a mitigated Coulomb attraction due to a vibrational contribution to the screening. Lattice screening is particularly relevant in ionic materials, where long-range optical phonon modes can lead to large electric fields~\cite{pavone}. Within the BSE formalism, these effects have been modeled by either accounting for additional screening based on simple approximations for the dielectric function~\cite{fuchs_08,bechstedt_ga203,Schleife2018,umari, umari_perovskites, highly_ionic,bokdam2016role} or from first principles via the inclusion of an additional contribution in the screened Coulomb interaction~\cite{filip_perovksites, alvertis_bse_exc_2023}. In either case, significant reductions of the excitonic binding energies have been reported, resulting in improved agreement with experimental results. However, these studies have thus far only examined the phonon-assisted effects on the lowest excitonic state, while the impact of lattice screening on higher-energy excitations and the overall absorption spectra have not yet been addressed.    

In this manuscript, we explore the influence of phonon-assisted screening on the excitation energies and spectral features for the polar semiconductors ZnS, GaN, and MgO from first principles. Our simulations consider all excitonic states and electron-hole transitions contributing to the optical absorption for energies up to 5~eV above the absorption edge. We separately account for  short-  and long-range screening effects at a fully {\it ab inito} level and further study the long-range effects within the commonly made Fr\"ohlich approximation. 

The manuscript is structured as follows: In Sec.~\ref{sec:theory}, we review the theoretical background for the inclusion of phonon-mediated screening effects within the BSE formalism. In Sec.~\ref{sec:exc_renorm}, we introduce a gauge-consistent computational workflow for the treatment of exciton-phonon interactions and discuss the computational parameters. In Sec.~\ref{sec:polar}, we apply this formalism to the polar semiconductors ZnS, GaN, and MgO. Conclusions and outlook are reported in Sec.~\ref{sec:conclusion}. 

\section{Theoretical background}\label{sec:theory}
In the following, we briefly outline the key steps required for the inclusion of phonon-assisted screening effects into the BSE formalism. The BSE can be written as an eigenvalue problem which, within the Tamm-Dancoff approximation, reads \cite{strinati1988application, rohlf_louie_2000}
\begin{align}\label{bse_eigen}
        \sum_{v'c'\k'}  \Bigl[&\left(\varepsilon_{c\k} - \varepsilon_{v\k}\right)\delta_{vv'}\,\delta_{cc'}\,\delta_{\k\k'} +  H^\text{x}_{vc\k,v'c'\k'} \nonumber\\
        & +   H^\text{dir}_{vc\k,v'c'\k'}(E^\lambda)  \Bigr] A^\lambda_{v'c'\k'} = E^\lambda A^\lambda_{vc\k}.
\end{align}
Here, $\lambda$ labels correlated electron-hole excitations with energies $E^\lambda$ and coupling coefficients $A^\lambda_{vc\k}$, which relate the exciton wavefunctions to the single-particle transitions via $|\Psi^\lambda\rangle = \sum_{vc\k}A^\lambda_{vc\k}|vc\k\rangle $. $\varepsilon_{v\k}$ and  $\varepsilon_{c\k}$ denote the quasi-particle energies at a given $\k$-point of valence and conduction states, respectively. The first term on the left-hand side in Eq.\;\eqref{bse_eigen} stems from independent quasi-particle transitions, and the exchange term $H^\text{x}$ contains the bare Coulomb interaction. The matrix elements of the direct term $H^\text{dir}$ are computed as~\cite{strinati1988application,rohlf_louie_2000}:
\begin{equation}\label{screen_dyn_real}
\begin{aligned}
   H^\text{dir}_{vc\k,v'c'\k'}(E^\lambda) &=   \langle vc\k |
       \frac{i}{2\pi}\int\text{\!\!d}\omega\, e^{-i\omega\eta}\, W(\r,\r',\omega) \times \\ 
       &\Biggl[\frac{1}{  E^\lambda - \left(\varepsilon_{c\k} - \varepsilon_{v'\mathbf{k'}}\right)  - \omega + i\eta } \\ & + \frac{1}{ E^\lambda - \left(\varepsilon_{c'\k'} - \varepsilon_{v\k}\right) + \omega +  i\eta}\Biggr] | v'c'\k' \rangle,
\end{aligned}  
\end{equation}
where $W$ is the total screened Coulomb potential and $\eta$ is a positive infinitesimal. Due to the dependence of Eq.~\eqref{screen_dyn_real} on $E^\lambda$, the BSE constitutes in principle a self-consistent problem.

\begin{figure*}[t]
    \centering
    \includegraphics[width=\linewidth]{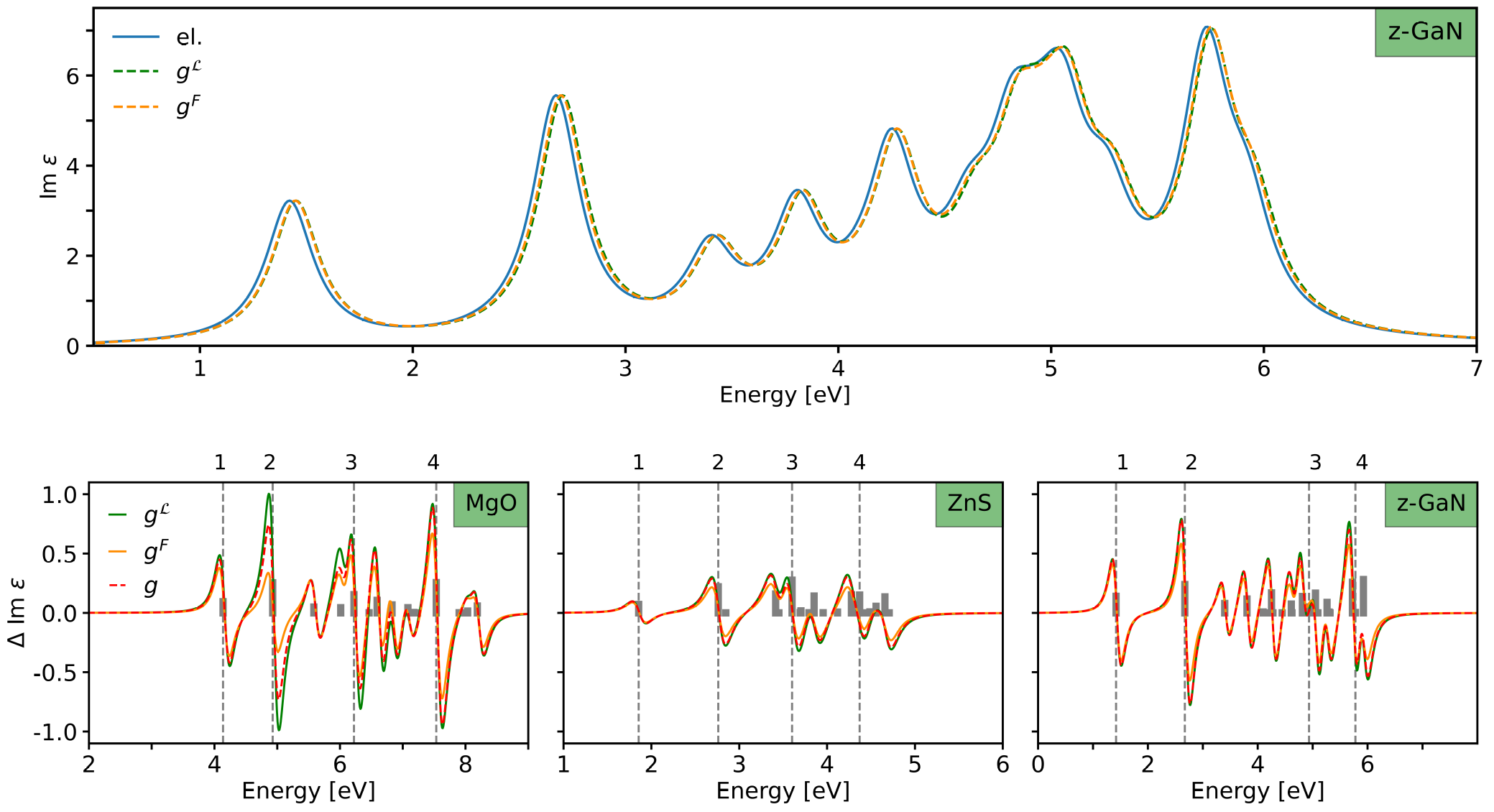}    
    \caption{\textbf{Phonon screening corrections to the absorption spectra. }Top: Imaginary part of the dielectric function (Eq.~\eqref{eq:im_diel}) of GaN from the BSE with pure electronic screening (blue) and with the phonon-screening corrected energies $\tilde{E}^\lambda = E^\lambda +  \Delta E^\lambda$ obtained from the long-range {\it ab initio} vertex $g^\mathcal{L}$ (green, Eq.~\eqref{eq:g_long}) and from the Fr\"ohlich model $g^F$ (yellow, Eq.~\eqref{eq:g_frohlich}). Bottom: Differences, $\Delta \Im \varepsilon = \Im\varepsilon_{\rm el} - \Im\varepsilon_{\rm el+ph}$, appearing due to phonon screening effects for MgO (left), ZnS (middle), and GaN (right). Gray bars indicate the oscillator strengths of all excitations, vertical dashed lines correspond to the excitation energies of the four most pronounced peaks. A broadening of 120\,meV is applied to mimic excitonic lifetimes. \label{fig:results_spectrum}  }
\end{figure*}

The starting point for the inclusion of lattice screening is the partitioning of the screened Coulomb interaction into electronic and vibrational contributions, \ie $W=W^{\rm el } + W^{\rm ph}$. The direct interaction term in the BSE Hamiltonian can be partitioned accordingly as $H^\text{dir} = H^\text{el} + H^\text{ph}$, where explicit expressions for the representation of the direct interaction  $H^\text{el}$ can be found elsewhere~\cite{rohlf_louie_2000,berkeleygw,exciting}. Most implementations of the BSE approach are typically tailored to capture the static screening ($\omega=0$) of the electronic contribution, while only recently the studies of excitonic properties have been extended to consider the effects of phonons on the excitonic screening and the exciton binding energies~\cite{filip_perovksites,alvertis_bse_exc_2023,umari, cudazzo_2020, antonius_exc_ph}.

Following Refs.~\cite{Giustino, filip_perovksites}, the phonon contribution to the screened Coulomb interaction can be written as
\begin{align}\label{eq:w_ph_real}
 W^\text{ph}(\r_1,\r_2,\omega) & = N^{-1}_p\,\sum_{\q\nu} g^*_{\q\nu}(\r_2)\,g_{\q\nu}(\r_1)\,D^\text{A}_{\q\nu}(\omega),
\end{align}
where $N_p$ is the total number of points in the $\q$-grid and $g_{\q\nu}(\r)=\Delta_{\q\nu}V(\r)$ the electron-phonon coupling potential, which is given by the change of the Kohn-Sham potential due to a phonon mode $\nu$ with frequency $\omega_{\q\nu}$ and crystal momentum $\q$. $D^\text{A}_{\q\nu}(\omega) = (\omega - \omega_{\q\nu} + i\eta)^{-1} -  (\omega + \omega_{\q\nu} - i\eta)^{-1}$ is the adiabatic phonon propagator. Inserting Eq.~\eqref{eq:w_ph_real} into Eq.~\eqref{screen_dyn_real} and making a unitary transformation to the exciton basis yields~\cite{filip_perovksites,alvertis_bse_exc_2023}
\begin{equation}\label{eq:h_dir_ph_exc_basis}
\begin{aligned}
      H^\text{ph}_{\lambda\lambda'}  =   - &\sum_{vc\k v'c'\k',\nu}A^{\lambda*}_{vc\k}A^{\lambda'}_{v'c'\k'} g_{cc'\nu}(\k',\q)g^*_{vv'\nu}(\k',\q) \\
      & \times \Biggl[\frac{1}{  E^\lambda - \left(\varepsilon_{c\k} - \varepsilon_{v'\mathbf{k'}}\right)  - \omega_{\q\nu} + i\eta } \\
    & + \frac{1}{ E^\lambda - \left(\varepsilon_{c'\k'} - \varepsilon_{v\k}\right)- \omega_{\q\nu} +  i\eta}\Biggr],
\end{aligned}
\end{equation}
with the electron phonon matrix elements $g_{mn\nu}(\k',\q)=\langle m\k'+\q|g_{\q\nu}|n\k'\rangle$ and $\q = \k - \k'$. This expression enables to compute the correction to the excitation energies $E^\lambda$ due to the phonon-assisted screening of the electron-hole interaction at first order perturbation theory as \cite{filip_perovksites}
\begin{equation}\label{eq:re_exc_corr}
\Delta E^\lambda = \text{Re} H^\text{ph}_{\lambda\lambda}.
\end{equation}

In order to disentangle effects of long-range and short-range interactions, we follow \cite{gius_verdi} and express the electron-phonon matrix elements as $g=g^\mathcal{S}+ g^\mathcal{L}$, where the long-range contribution $g^\mathcal{L}$ can be obtained from first principles as 
\begin{equation}\label{eq:g_long}
\begin{aligned}
g^\mathcal{L}_{mn\nu}(\k,\q) & = i\,\frac{4\pi}{\Omega^{\phantom{e}}_\text{UC}}\sum_\kappa\left( \frac{1}{2M_\kappa\omega_{\q\nu}}   \right)^{\!\!1/2} \\[10pt]
    & \quad \times \sum_{\G\neq-\q} \frac{(\q+\G)\cdot\mathbf{Z}^*_\kappa\cdot\mathbf{e}_{\kappa\nu}(\q)}{|\q+\G|\cdot\boldsymbol{\varepsilon}_\infty\cdot|\q+\G} \\[10pt]
      & \quad \times \langle m\k+\q|e^{i\mathbf{(q+G)\cdot r}}|n\k\rangle e^{-i\mathbf{(q+G)\cdot\boldsymbol{\tau}_{\!\kappa}}}.
      \end{aligned}
\end{equation}
In this expression, the unit cell volume is denoted as $\Omega_{\rm UC}$, $\mathbf{G}$ is a reciprocal lattice vector, and $\boldsymbol{\varepsilon}_\infty$ is the high-frequency dielectric tensor. Each atom labeled by $\kappa$ is characterized by the Born effective charge tensor $\mathbf{Z}^*_\kappa$, mass $M_\kappa$, phonon eigenvectors $\mathbf{e}_{\kappa\nu}$, and the atomic position $\boldsymbol{\tau}_{\!\kappa}$.
In our work, all matrix elements $g$ are obtained from density-functional perturbation theory (DFPT)~\cite{baroni_dfpt_2001,giannozzi2005density,kouba2001}. 

For the long-range coupling, we further evaluate the effects of the simplified Fr\"ohlich model~\cite{froehlich}. Here, it is assumed that the coupling is dominated by long-range interactions with a dispersionless longitudinal optical phonon (LO) mode with frequency $\omega_{\rm LO}$, such that the matrix elements can be approximated as
\begin{equation}\label{eq:g_frohlich}
    g^{\rm F}(\q) = \frac{i}{|\q|}\left[ \frac{4\pi}{\Omega_{\rm UC}}\frac{\omega_{\rm LO}}{2}\left(\frac{1}{\varepsilon_\infty}  - \frac{1}{\varepsilon_0} \right)\right]^{1/2}.
\end{equation}
$\varepsilon_\infty$ and $\varepsilon_0$ are the high-frequency and the static dielectric constants, respectively.

\section{Implementation and computational details}\label{sec:exc_renorm}
We have implemented Eqs.~\eqref{eq:h_dir_ph_exc_basis} and \eqref{eq:re_exc_corr} in the all-electron code \exciting{} \cite{exciting} to assess the influence of lattice screening on excitons. In the following, we outline the computational procedure adopted in our study. First, a groundstate DFT calculation is performed to obtain the single-particle Kohn-Sham wavefunctions and eigenvalues. These quantities serve as input for the construction of the BSE Hamiltonian. Subsequently, the electron-phonon matrix elements are computed  from DFPT. We do not make use of Wannier-Fourier interpolation, since this step leads to inconsistencies between the electron-phonon matrix elements and the exciton coupling coefficients \cite{giustino_eph_wannier_2007,alvertis_bse_exc_2023}. To ensure gauge consistency in the calculation of the long-range part (Eq.~\eqref{eq:g_long}), the exact same dipole matrix elements are used as in the construction of the BSE Hamiltonian. An alternative strategy to the one adopted here consists in first computing the electronic wave-functions based on Wannier-interpolation, which then serve as input for both the electron-phonon matrix elements and the BSE envelope functions \cite{alvertis_bse_exc_2023}. Both are then used to compute the corrections to the excitation energies (Eq.~\eqref{eq:h_dir_ph_exc_basis}), which we solve in a one-shot procedure, \ie without updating the corrected energies $E^\lambda$.

The insertion of the long-range matrix element (Eq.~\eqref{eq:g_long}) into Eq.~\eqref{eq:h_dir_ph_exc_basis} leads to an integrable divergence that, however, requires special treatment for numerical stability. To deal with this divergence, we follow the strategy already introduced and applied to the Coulomb divergences in $GW$ calculations \cite{hybrtsen_louie_1986, puschnig_draxl_2002}. More specifically, we use a spherical averaging scheme that performs well for cubic or only weakly anisotropic materials like those considered here. The product of two matrix elements is approximated in the limit of $\q\rightarrow0$ as 
\begin{equation}
    g(\k,\q\rightarrow0)g^*(\k,\q\rightarrow0)\approx \frac{\tilde{g}(\k,\q\rightarrow 0)\tilde{g}^*(\k,\q\rightarrow 0)}{|\q|^2},
\end{equation}
where $\tilde{g}$ is the non-divergent contribution of the matrix elements. This expression allows us to integrate out the divergence by averaging the product in a finite subcell around $\q=0$ according to
\begin{align}\label{eq:av_g}
    g(\k,\q\rightarrow0) & g^*(\k,\q\rightarrow0)\approx  \\ & \tilde{g}(\k,\q\rightarrow 0)\tilde{g}^*(\k,\q\rightarrow 0)  \nonumber  \frac{1}{V_q}\int d\q \frac{1}{|\q|^2}, 
\end{align}
where $V_q=\Omega_{\rm BZ} / N_\k$ is the volume of the subcell defined by the Brillouin zone volume $\Omega_{\rm BZ}$.

 All calculations are carried out with the \exciting\ \cite{exciting} code, which employs the all-electron full-potential (linearized) augmented planewave plus local orbital [(L)APW+lo] method. We use the PBEsol parametrization \cite{pbe_sol} of the generalized-gradient approximation for exchange-correlation effects. The basis-set size is determined in terms of the muffin-tin radius $R_\text{MT}$  by the cut-off $R_\text{MT}|\G+\k|_\text{max}$\,=\,$7$, and an $8\!\times\!8\!\times\!8$ $\k$-point grid is used for all studied materials. Phonons are computed from DFPT on the same $\q$-grid as the groundstate.  Excitonic properties are computed with the BSE package of \exciting\ \cite{Vorwerk_2019}, where 100 empty states are used for the calculation of the RPA screening. Local-field effects are included up to $|\G+\q|_\text{max}\!=\!3.0\,\text{a.u.}^{-1}$. In the setup of the BSE Hamiltonian, 4 conduction and 5 valence bands are considered, and a Lorentzian broadening of 120\,meV is employed to mimic lifetime effects. The $\k/\q$-point grids used here are the same as for the groundstate.

 The momentum grids employed in our work uniformly cover the whole Brillouin zone. This is required for the study of high-energy excitations, but has the disadvantage that exciton binding energies of the lowest excitations, that are typically highly localized around $\Gamma$, may be hard to converge \cite{alvertis_nonuniform_bz_23}. In this work, however, we are interested in the changes of excitation energies upon lattice screening, $\Delta E^\lambda$, rather than their absolute values, for which we find comparable values to previously published results despite the use of coarser grids (see Section~\ref{sec:polar}). 

\section{Phonon-screening effects in the polar semiconductors Z\lowercase{n}s, G\lowercase{a}N, and M\lowercase{g}O}\label{sec:polar}

We apply the developed approach to polar semiconductors, and choose MgO,  ZnS, and GaN in the cubic zincblende structure (z-GaN) as the prototypical examples. For all materials, we compute the electron-phonon corrections to the first 200 excitonic states. 

We first study the effects of phonon screening on the optical absorption spectra, for which we set the parameter $\eta$ (see Eq.~\eqref{eq:h_dir_ph_exc_basis}) to 50\,meV. From the solution of the BSE, the imaginary part of the dielectric function is computed as
\begin{equation}\label{eq:im_diel}
\Im \varepsilon(\omega) = \frac{4\pi^2}{V}\sum_\lambda |\hat{\mathbf{e}}\cdot\mathbf{t}^\lambda|^2 \delta (E^\lambda - \omega),
\end{equation}
where $\mathbf{t}^\lambda$ are the excitonic oscillator strengths, $\hat{\mathbf{e}}$ is the polarization direction of the light, and $V$ is the crystal volume. Accordingly, the effects of phonon screening on the optical spectrum can be evaluated by inserting the corrected energies $\tilde{E}^\lambda = E^\lambda +  \Delta E^\lambda$ in the evaluation of Eq.~\eqref{eq:im_diel}. The top panel of Fig.~\ref{fig:results_spectrum} shows the spectrum obtained for GaN with pure electronic screening and including electron-phonon corrections using the {\it ab initio} long-range corrections and the Fr\"ohlich model, respectively. The high-frequency (static) dielectric constants used in the evaluation of the Fr\"ohlich model are determined from DFPT to be 3.0 (9.0) for MgO, 5.7 (8.6) for ZnS, and 6.0 (9.6) for GaN. The spectra obtained using the full {\it ab initio} matrix elements are found to be almost identical to the ones obtained using only the long-range parts and are therefore not shown. 

To resolve the small changes between the different spectra, the differences between the purely electronic and the phonon-corrected spectra are shown in the bottom panels for all investigated materials, with gray bars indicating the locations and oscillator strengths of the excitations. The excitation energies corresponding to the four most pronounced peaks are marked by dashed lines. For all three materials, we find that the phonon screening results only in small corrections to the optical spectra, given by almost rigid shifts of the peaks corresponding to the changes in the excitation energies.  Apart from the second peak in MgO, we further find that the differences obtained with the Fr\"ohlich model are almost identical to those obtained with the {\it ab initio} long-range vertex as well as with those obtained using the full vertex including short-range corrections. This observation indicates that the Fr\"ohlich model is a good approximation in the systems under study and that short-range corrections are negligible. These results are further corroborated by the observation that all materials fulfill the main condition required for the validity of the Fr\"ohlich model, namely, they exhibit only one polar LO phonon. 

\begin{table}
\caption{Renormalization of the excitation energies in meV for the four states with the highest oscillator strengths marked in the lower panel of Fig.~\ref{fig:results_spectrum}. The three columns per material correspond to the different approximations to compute the electron-phonon matrix elements.\label{tab:exc}}
\begin{tabular}{c| c c c|c c c|c c c} 
 \hline
 \hline
   &  & MgO &  &  & ZnS & & & z-GaN &\\ 
 \hline 
 Peak & $\Delta E$  & $\Delta E^\mathcal{L}$ & $\Delta E^{\rm F}$ & $\Delta E$  & $\Delta E^\mathcal{L}$ & $\Delta E^{\rm F}$ & $\Delta E$  & $\Delta E^\mathcal{L}$ & $\Delta E^{\rm F}$  \\ [0.5ex] 
 \hline
 1 & 50 & 53 & 42 & 12 & 12 & 11 & 30 & 30 & 27\\ 
 2 & 42 & 30 & 35 & 12 & 12 & 9 & 29 & 31 & 22\\
 3 & 63 & 52 & 41  & 13 & 14 & 9 & 32 &31 & 28\\
 4 &  40& 39 & 28 & 12 & 13 & 9 & 30 & 31 & 27 \\
 \hline
 \hline
\end{tabular}
\end{table}

\begin{figure}
    \centering
    \includegraphics[width=0.8\linewidth]{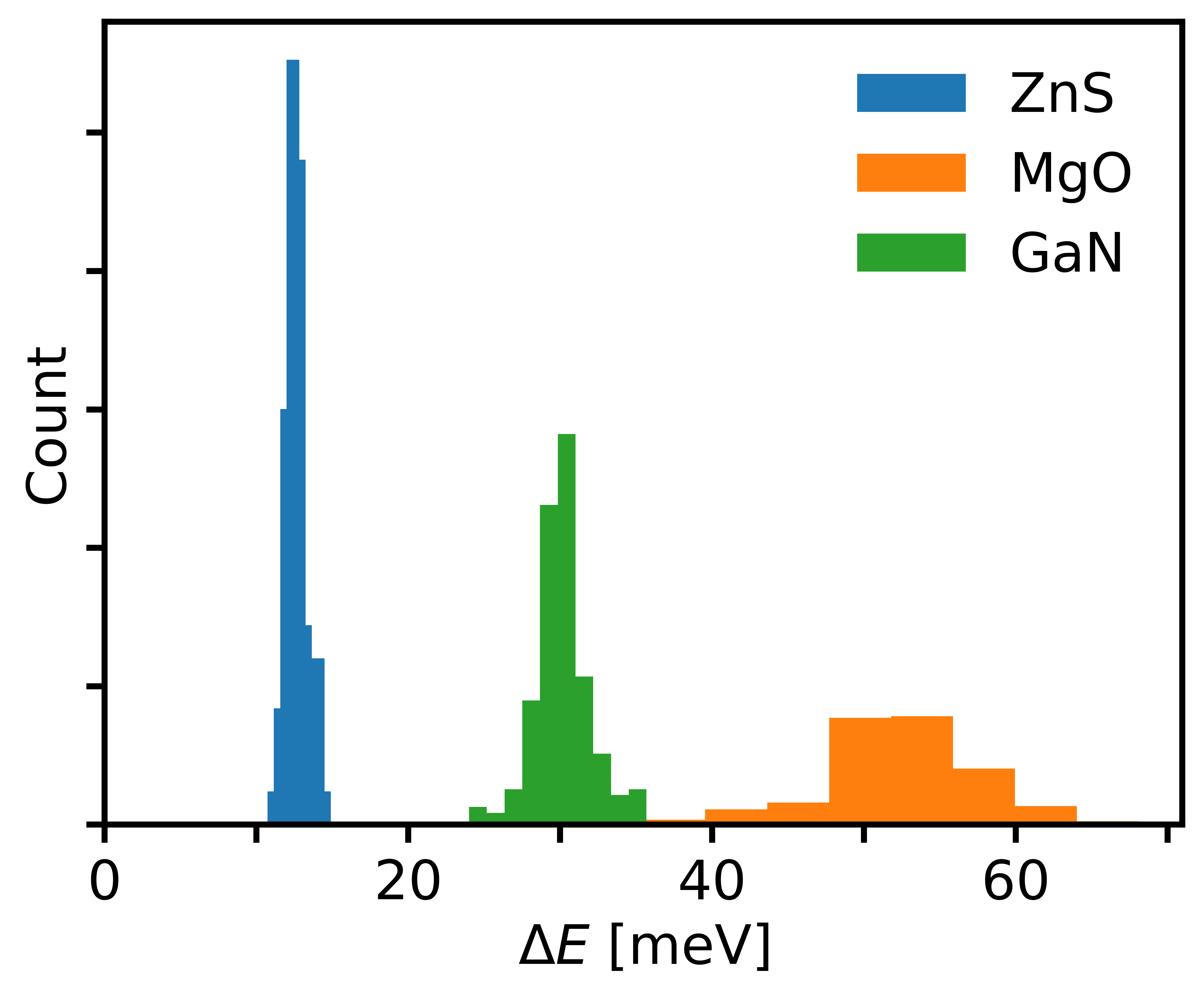}
    \caption{Normalized histograms of the corrections to the excitonic energies.}
    \label{fig:hist_corrections}
\end{figure}

We now proceed to study the effects of phonon screening on the individual excitations and consider the corrections to the excitation energies for the four states with the highest oscillator strengths marked in the lower row of Fig.~\ref{fig:results_spectrum}. Specifically, we evaluate  Eq.~\eqref{eq:h_dir_ph_exc_basis} and \eqref{eq:re_exc_corr}  for the full electron-phonon matrix element $g$ and  the  {\it ab initio} long-range contribution $g^{\mathcal{L}}$ (Eq.~\eqref{eq:g_long}) to obtain the corrections to the excitation energies $\Delta E$ and $\Delta E^{\mathcal{L}}$, respectively. In addition, we evaluate the effect of the Fr\"ohlich model (Eq.~\eqref{eq:g_frohlich}) and denote the resulting corrections as $\Delta E^{\rm F}$. Table~\ref{tab:exc} shows the corresponding results for $\Delta E$ for the four states of the three materials (Fig.~\ref{fig:results_spectrum}). For the excitonic groundstates (peak 1), the computed shifts in the excitation energy are in the order of few tens of meV for all states in all three materials, which is consistent which the corrections obtained for the groundstate in various other semiconductors~\cite{filip_perovksites,alvertis_bse_exc_2023}. For MgO, corrections larger than 50\,meV arise due to phonon corrections, whereas for GaN and ZnS only corrections of 27 and 12\,meV, respectively, are found. This is plausible considering that in MgO the ratio $\varepsilon_0/\varepsilon_\infty$ is the largest of all investigated systems. For higher excitations, similar corrections can be observed. To illustrate that this is a general trend, we show in Fig.~\ref{fig:hist_corrections} the normalized histograms of all computed corrections. The corrections for a given material are all comparable in magnitude with increasing spread for larger corrections. 
 
\begin{figure*}
\centering
\includegraphics[width=\textwidth]{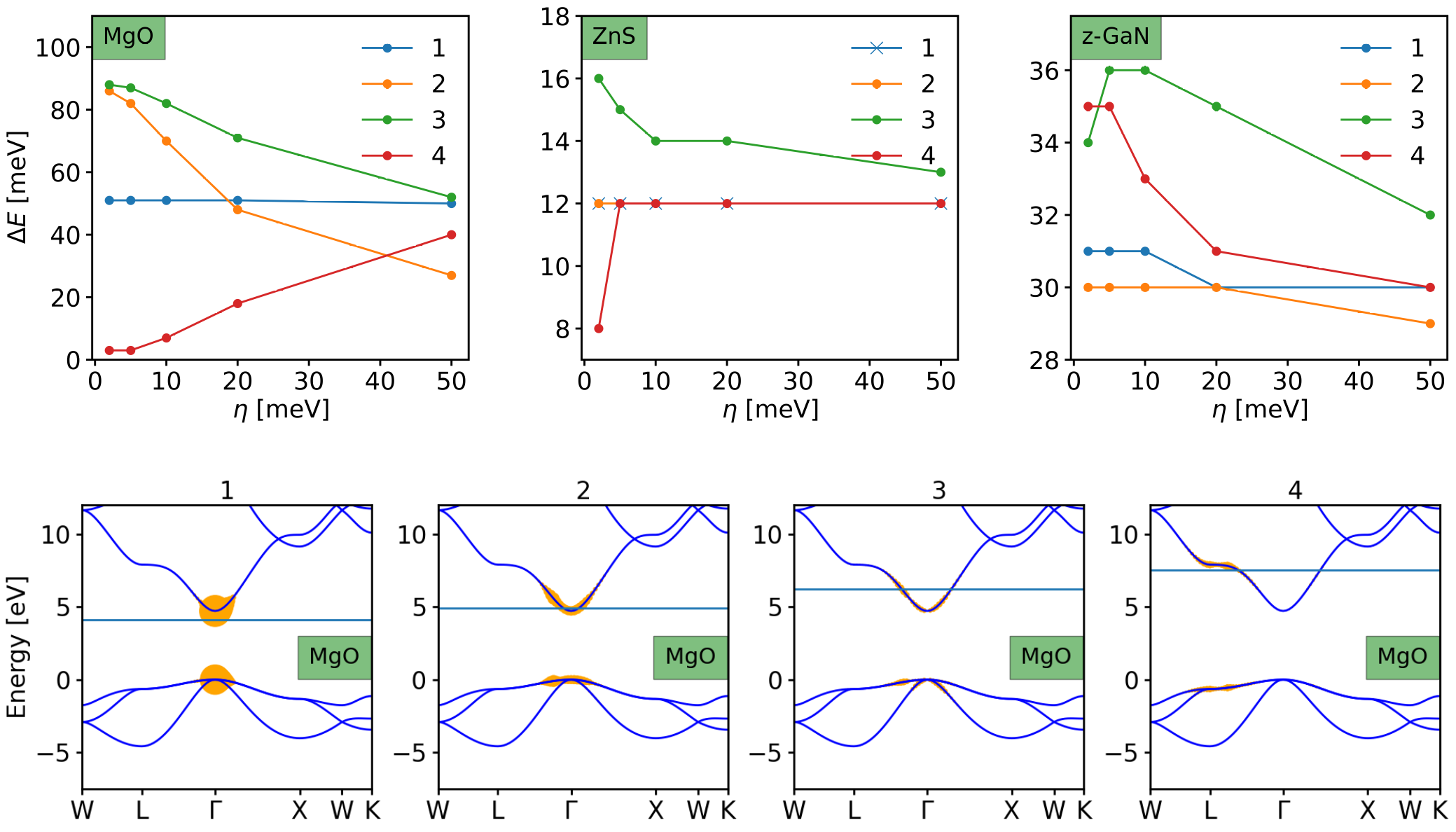}
\caption{Top row: Dependence of the phonon-assisted renormalization of the excitation energies on the parameter~$\eta$. Bottom row: Excitonic weights $w_{c\k}$ and $w_{v\k}$ (orange dots) for the four most prominent peaks in Fig.~\ref{fig:results_spectrum} in MgO superimposed on the bandstructure (blue lines). The blue horizontal lines indicates the excitation energies $E^\lambda$, with the valence band maximum set to zero. \label{fig:exc:_eta}}
\end{figure*}

Similar to the absorption spectrum, we find good agreement between the values obtained with the full vertex and the pure long-range vertex, indicating that also here short-ranged corrections are negligible. For all materials, we observe a good agreement of the Fr\"ohlich model with respect to the {\it ab initio} results, but for all materials the former slightly underestimates the {\it ab initio} results.  

In MgO, some larger discrepancies between the corrections obtained using the long-range contributions and the full matrix elements are found for the excitonic states 2 and 3, respectively, namely , 42 vs 30\, meV and 63 vs 52\,meV. We assume that this is due to their pronounced dependence on $\eta$, such that also small differences in the matrix elements can lead to visible effects. In order to systematically investigate the convergence with respect to $\eta$, the top row of Fig.~\ref{fig:exc:_eta} shows the corrections for the four states for $\eta \in [2,5,10,20,50]\,$meV computed using the full matrix elements $g$. While the corrections for some of the states have a very strong dependence on $\eta$, others do not. Specifically, we find that for all of the investigated materials, the corrections corresponding to the first peak, stemming from the excitonic groundstate, show only a very weak dependence on $\eta$, whereas the higher excitations show stronger dependences, the strongest being found for MgO. 

To understand this situation better, it is helpful to introduce the excitonic weights of conduction and valence bands as $ w_{c\k} = \sum_v |A_{vc\k}|^2$ and $w_{v\k}  = \sum_c |A_{vc\k}|^2$, respectively. Exemplary for the three materials, the bottom row of Fig.~\ref{fig:exc:_eta} shows the results for the four selected peaks of MgO. Only for the excitonic groundstate (peak 1), $E^\lambda < \varepsilon_{c\k} - \varepsilon_{v\k}$ over the whole Brillouin zone, while the other states have non-vanishing coupling coefficients $A^\lambda_{vc\k}$ for both $E^\lambda < \varepsilon_{c\k} - \varepsilon_{v\k}$ and $E^\lambda > \varepsilon_{c\k} - \varepsilon_{v\k}$. This results in the fact that the higher excitations can have contributions, where $E^\lambda - (\varepsilon_{c\k} - \varepsilon_{v\k}) -\omega_{\q\nu}$ is very small and therefore $\Delta E$ diverges for $\eta=0$, since the contribution of a given $\k$-point to the phonon corrections $\Delta E^\lambda$ is proportional to $ (E^\lambda - \left(\varepsilon_{c\k} - \varepsilon_{v'\mathbf{k'}}\right)  - \omega_{\q\nu} + i\eta)^{-1}$ (see Eq.~\eqref{eq:h_dir_ph_exc_basis}). Such divergence can be resolved using $\eta>0$ but, depending on the number of $\k$-points contributing with large values, can result in the strong dependence on $\eta$ as seen in the top row of Fig.~\ref{fig:exc:_eta}. Similar results are obtained for ZnS and GaN. 
 
We note that the corrections computed in this work do only take into account the effects of lattice screening on the electron-hole interaction, but do not include effects of lattice distortions due the exciton formation. The latter would require the inclusion of exciton-phonon interactions~\cite{yang_exc_struc_relax_22,giustino_exc_polarons_24} which is out of the scope of the presented framework. 

\section{Conclusions}\label{sec:conclusion}
In conclusion, we presented an  {\it ab initio} study of the effects of electron-phonon interaction on the screening of excitons. We investigated the phonon-assisted renormalization of the excitation energies and optical properties for the prototypical polar semiconductors MgO, ZnS, and GaN. We find  renormalizations of the excitation energies on the order of tens of meV across all excitonic states. By accounting for the phonon-assisted screening effects for the lowest 200 excitonic states, we further explore the impact of electron-phonon interaction on the full optical absorption spectra. We find that vibrational screening can renormalize the main absorption peaks by up to 50~meV. These effects are not specific to bound excitons, but obvioulsy also apply to unbound electron-hole pairs. Overall, the lattice contribution to the exciton screening is largely dominated by the long-range Fröhlich coupling involving polar longitudinal-optical phonons. Conversely, short-range effects are found to be negligible. We thus expect vibrational screening to be insignificant for the excitonic properties of non-polar and weakly polar insulators. Further, we showed that the Fr\"ohlich model serves as a good approximation for the study of phonon screening effects on the absorption spectra, which is straightforward to include in existing implementations of the BSE formalism.

\section*{Acknowledgements}
Funding from the Deutsche Forschungsgemeinschaft (DFG), projects 182087777, 499426961, and 434434223 is acknowledged. We thank Ignacio Oliva Gonzalez and Sebastian Tillack for insightful discussions and for providing the DFPT code for the calculation of the electron-phonon matrix elements. \\

\phantom{.}\bibliographystyle{apsrev4-1}
\bibliography{article}

\providecommand{\noopsort}[1]{}\providecommand{\singleletter}[1]{#1}%
\begin{thebibliography}{71}%
\makeatletter
\providecommand \@ifxundefined [1]{%
 \@ifx{#1\undefined}
}%
\providecommand \@ifnum [1]{%
 \ifnum #1\expandafter \@firstoftwo
 \else \expandafter \@secondoftwo
 \fi
}%
\providecommand \@ifx [1]{%
 \ifx #1\expandafter \@firstoftwo
 \else \expandafter \@secondoftwo
 \fi
}%
\providecommand \natexlab [1]{#1}%
\providecommand \enquote  [1]{``#1''}%
\providecommand \bibnamefont  [1]{#1}%
\providecommand \bibfnamefont [1]{#1}%
\providecommand \citenamefont [1]{#1}%
\providecommand \href@noop [0]{\@secondoftwo}%
\providecommand \href [0]{\begingroup \@sanitize@url \@href}%
\providecommand \@href[1]{\@@startlink{#1}\@@href}%
\providecommand \@@href[1]{\endgroup#1\@@endlink}%
\providecommand \@sanitize@url [0]{\catcode `\\12\catcode `\$12\catcode
  `\&12\catcode `\#12\catcode `\^12\catcode `\_12\catcode `\%12\relax}%
\providecommand \@@startlink[1]{}%
\providecommand \@@endlink[0]{}%
\providecommand \url  [0]{\begingroup\@sanitize@url \@url }%
\providecommand \@url [1]{\endgroup\@href {#1}{\urlprefix }}%
\providecommand \urlprefix  [0]{URL }%
\providecommand \Eprint [0]{\href }%
\providecommand \doibase [0]{http://dx.doi.org/}%
\providecommand \selectlanguage [0]{\@gobble}%
\providecommand \bibinfo  [0]{\@secondoftwo}%
\providecommand \bibfield  [0]{\@secondoftwo}%
\providecommand \translation [1]{[#1]}%
\providecommand \BibitemOpen [0]{}%
\providecommand \bibitemStop [0]{}%
\providecommand \bibitemNoStop [0]{.\EOS\space}%
\providecommand \EOS [0]{\spacefactor3000\relax}%
\providecommand \BibitemShut  [1]{\csname bibitem#1\endcsname}%
\let\auto@bib@innerbib\@empty
\bibitem [{\citenamefont {Yu}\ and\ \citenamefont
  {Cardona}(2005)}]{cardona2005fundamentals}%
  \BibitemOpen
  \bibfield  {author} {\bibinfo {author} {\bibfnamefont {P.~Y.}\ \bibnamefont
  {Yu}}\ and\ \bibinfo {author} {\bibfnamefont {M.}~\bibnamefont {Cardona}},\
  }\href {\doibase 10.1007/978-3-642-00710-1} {\emph {\bibinfo {title}
  {Fundamentals of semiconductors}}}\ (\bibinfo  {publisher} {Springer},\
  \bibinfo {address} {Berlin},\ \bibinfo {year} {2005})\BibitemShut {NoStop}%
\bibitem [{\citenamefont {Mueller}\ and\ \citenamefont
  {Malic}(2018)}]{Mueller2018-tn}%
  \BibitemOpen
  \bibfield  {author} {\bibinfo {author} {\bibfnamefont {T.}~\bibnamefont
  {Mueller}}\ and\ \bibinfo {author} {\bibfnamefont {E.}~\bibnamefont
  {Malic}},\ }\href@noop {} {\bibfield  {journal} {\bibinfo  {journal} {Npj 2D
  Mater. Appl.}\ }\textbf {\bibinfo {volume} {2}} (\bibinfo {year}
  {2018})}\BibitemShut {NoStop}%
\bibitem [{\citenamefont {Li}\ \emph {et~al.}(2023)\citenamefont {Li},
  \citenamefont {Yan},\ and\ \citenamefont {Fang}}]{Li2023-sw}%
  \BibitemOpen
  \bibfield  {author} {\bibinfo {author} {\bibfnamefont {Z.}~\bibnamefont
  {Li}}, \bibinfo {author} {\bibfnamefont {T.}~\bibnamefont {Yan}}, \ and\
  \bibinfo {author} {\bibfnamefont {X.}~\bibnamefont {Fang}},\ }\href@noop {}
  {\bibfield  {journal} {\bibinfo  {journal} {Nat. Rev. Mater.}\ }\textbf
  {\bibinfo {volume} {8}},\ \bibinfo {pages} {587} (\bibinfo {year}
  {2023})}\BibitemShut {NoStop}%
\bibitem [{\citenamefont {Tang}\ \emph {et~al.}(2022)\citenamefont {Tang},
  \citenamefont {Li}, \citenamefont {Jiang}, \citenamefont {Gao}, \citenamefont
  {Li}, \citenamefont {Li}, \citenamefont {Huang}, \citenamefont {Kang},
  \citenamefont {Wang},\ and\ \citenamefont {Zhang}}]{Tang2022-un}%
  \BibitemOpen
  \bibfield  {author} {\bibinfo {author} {\bibfnamefont {R.}~\bibnamefont
  {Tang}}, \bibinfo {author} {\bibfnamefont {G.}~\bibnamefont {Li}}, \bibinfo
  {author} {\bibfnamefont {Y.}~\bibnamefont {Jiang}}, \bibinfo {author}
  {\bibfnamefont {N.}~\bibnamefont {Gao}}, \bibinfo {author} {\bibfnamefont
  {J.}~\bibnamefont {Li}}, \bibinfo {author} {\bibfnamefont {C.}~\bibnamefont
  {Li}}, \bibinfo {author} {\bibfnamefont {K.}~\bibnamefont {Huang}}, \bibinfo
  {author} {\bibfnamefont {J.}~\bibnamefont {Kang}}, \bibinfo {author}
  {\bibfnamefont {T.}~\bibnamefont {Wang}}, \ and\ \bibinfo {author}
  {\bibfnamefont {R.}~\bibnamefont {Zhang}},\ }\href@noop {} {\bibfield
  {journal} {\bibinfo  {journal} {ACS Appl. Electron. Mater.}\ }\textbf
  {\bibinfo {volume} {4}},\ \bibinfo {pages} {188} (\bibinfo {year}
  {2022})}\BibitemShut {NoStop}%
\bibitem [{\citenamefont {Udagawa}\ \emph {et~al.}(2014)\citenamefont
  {Udagawa}, \citenamefont {Sasabe}, \citenamefont {Cai},\ and\ \citenamefont
  {Kido}}]{Udagawa2014}%
  \BibitemOpen
  \bibfield  {author} {\bibinfo {author} {\bibfnamefont {K.}~\bibnamefont
  {Udagawa}}, \bibinfo {author} {\bibfnamefont {H.}~\bibnamefont {Sasabe}},
  \bibinfo {author} {\bibfnamefont {C.}~\bibnamefont {Cai}}, \ and\ \bibinfo
  {author} {\bibfnamefont {J.}~\bibnamefont {Kido}},\ }\href {\doibase
  10.1002/adma.201401621} {\bibfield  {journal} {\bibinfo  {journal} {Advanced
  Materials}\ }\textbf {\bibinfo {volume} {26}},\ \bibinfo {pages}
  {5062–5066} (\bibinfo {year} {2014})}\BibitemShut {NoStop}%
\bibitem [{\citenamefont {Sundaram}\ \emph {et~al.}(2013)\citenamefont
  {Sundaram}, \citenamefont {Engel}, \citenamefont {Lombardo}, \citenamefont
  {Krupke}, \citenamefont {Ferrari}, \citenamefont {Avouris},\ and\
  \citenamefont {Steiner}}]{Sundaram2013-ac}%
  \BibitemOpen
  \bibfield  {author} {\bibinfo {author} {\bibfnamefont {R.~S.}\ \bibnamefont
  {Sundaram}}, \bibinfo {author} {\bibfnamefont {M.}~\bibnamefont {Engel}},
  \bibinfo {author} {\bibfnamefont {A.}~\bibnamefont {Lombardo}}, \bibinfo
  {author} {\bibfnamefont {R.}~\bibnamefont {Krupke}}, \bibinfo {author}
  {\bibfnamefont {A.~C.}\ \bibnamefont {Ferrari}}, \bibinfo {author}
  {\bibfnamefont {P.}~\bibnamefont {Avouris}}, \ and\ \bibinfo {author}
  {\bibfnamefont {M.}~\bibnamefont {Steiner}},\ }\href@noop {} {\bibfield
  {journal} {\bibinfo  {journal} {Nano Lett.}\ }\textbf {\bibinfo {volume}
  {13}},\ \bibinfo {pages} {1416} (\bibinfo {year} {2013})}\BibitemShut
  {NoStop}%
\bibitem [{\citenamefont {Grancini}\ \emph {et~al.}(2012)\citenamefont
  {Grancini}, \citenamefont {Maiuri}, \citenamefont {Fazzi}, \citenamefont
  {Petrozza}, \citenamefont {Egelhaaf}, \citenamefont {Brida}, \citenamefont
  {Cerullo},\ and\ \citenamefont {Lanzani}}]{Grancini2012}%
  \BibitemOpen
  \bibfield  {author} {\bibinfo {author} {\bibfnamefont {G.}~\bibnamefont
  {Grancini}}, \bibinfo {author} {\bibfnamefont {M.}~\bibnamefont {Maiuri}},
  \bibinfo {author} {\bibfnamefont {D.}~\bibnamefont {Fazzi}}, \bibinfo
  {author} {\bibfnamefont {A.}~\bibnamefont {Petrozza}}, \bibinfo {author}
  {\bibfnamefont {H.-J.}\ \bibnamefont {Egelhaaf}}, \bibinfo {author}
  {\bibfnamefont {D.}~\bibnamefont {Brida}}, \bibinfo {author} {\bibfnamefont
  {G.}~\bibnamefont {Cerullo}}, \ and\ \bibinfo {author} {\bibfnamefont
  {G.}~\bibnamefont {Lanzani}},\ }\href {\doibase 10.1038/nmat3502} {\bibfield
  {journal} {\bibinfo  {journal} {Nature Materials}\ }\textbf {\bibinfo
  {volume} {12}},\ \bibinfo {pages} {29–33} (\bibinfo {year}
  {2012})}\BibitemShut {NoStop}%
\bibitem [{\citenamefont {Kojima}\ \emph {et~al.}(2009)\citenamefont {Kojima},
  \citenamefont {Teshima}, \citenamefont {Shirai},\ and\ \citenamefont
  {Miyasaka}}]{Kojima2009}%
  \BibitemOpen
  \bibfield  {author} {\bibinfo {author} {\bibfnamefont {A.}~\bibnamefont
  {Kojima}}, \bibinfo {author} {\bibfnamefont {K.}~\bibnamefont {Teshima}},
  \bibinfo {author} {\bibfnamefont {Y.}~\bibnamefont {Shirai}}, \ and\ \bibinfo
  {author} {\bibfnamefont {T.}~\bibnamefont {Miyasaka}},\ }\href {\doibase
  10.1021/ja809598r} {\bibfield  {journal} {\bibinfo  {journal} {Journal of the
  American Chemical Society}\ }\textbf {\bibinfo {volume} {131}},\ \bibinfo
  {pages} {6050–6051} (\bibinfo {year} {2009})}\BibitemShut {NoStop}%
\bibitem [{\citenamefont {Malinkiewicz}\ \emph {et~al.}(2013)\citenamefont
  {Malinkiewicz}, \citenamefont {Yella}, \citenamefont {Lee}, \citenamefont
  {Espallargas}, \citenamefont {Graetzel}, \citenamefont {Nazeeruddin},\ and\
  \citenamefont {Bolink}}]{Malinkiewicz2013}%
  \BibitemOpen
  \bibfield  {author} {\bibinfo {author} {\bibfnamefont {O.}~\bibnamefont
  {Malinkiewicz}}, \bibinfo {author} {\bibfnamefont {A.}~\bibnamefont {Yella}},
  \bibinfo {author} {\bibfnamefont {Y.~H.}\ \bibnamefont {Lee}}, \bibinfo
  {author} {\bibfnamefont {G.~M.}\ \bibnamefont {Espallargas}}, \bibinfo
  {author} {\bibfnamefont {M.}~\bibnamefont {Graetzel}}, \bibinfo {author}
  {\bibfnamefont {M.~K.}\ \bibnamefont {Nazeeruddin}}, \ and\ \bibinfo {author}
  {\bibfnamefont {H.~J.}\ \bibnamefont {Bolink}},\ }\href {\doibase
  10.1038/nphoton.2013.341} {\bibfield  {journal} {\bibinfo  {journal} {Nature
  Photonics}\ }\textbf {\bibinfo {volume} {8}},\ \bibinfo {pages} {128–132}
  (\bibinfo {year} {2013})}\BibitemShut {NoStop}%
\bibitem [{\citenamefont {Stranks}\ and\ \citenamefont
  {Snaith}(2015)}]{Stranks2015-ne}%
  \BibitemOpen
  \bibfield  {author} {\bibinfo {author} {\bibfnamefont {S.~D.}\ \bibnamefont
  {Stranks}}\ and\ \bibinfo {author} {\bibfnamefont {H.~J.}\ \bibnamefont
  {Snaith}},\ }\href@noop {} {\bibfield  {journal} {\bibinfo  {journal} {Nat.
  Nanotechnol.}\ }\textbf {\bibinfo {volume} {10}},\ \bibinfo {pages} {391}
  (\bibinfo {year} {2015})}\BibitemShut {NoStop}%
\bibitem [{\citenamefont {Kohn}\ and\ \citenamefont {Sham}(1965)}]{ks_dft_65}%
  \BibitemOpen
  \bibfield  {author} {\bibinfo {author} {\bibfnamefont {W.}~\bibnamefont
  {Kohn}}\ and\ \bibinfo {author} {\bibfnamefont {L.~J.}\ \bibnamefont
  {Sham}},\ }\href {\doibase 10.1103/PhysRev.140.A1133} {\bibfield  {journal}
  {\bibinfo  {journal} {Phys. Rev.}\ }\textbf {\bibinfo {volume} {140}},\
  \bibinfo {pages} {A1133} (\bibinfo {year} {1965})}\BibitemShut {NoStop}%
\bibitem [{\citenamefont {Hybertsen}\ and\ \citenamefont
  {Louie}(1986{\natexlab{a}})}]{hybertsen1986electron}%
  \BibitemOpen
  \bibfield  {author} {\bibinfo {author} {\bibfnamefont {M.~S.}\ \bibnamefont
  {Hybertsen}}\ and\ \bibinfo {author} {\bibfnamefont {S.~G.}\ \bibnamefont
  {Louie}},\ }\href {\doibase 10.1103/PhysRevB.34.5390} {\bibfield  {journal}
  {\bibinfo  {journal} {Phys. Rev. B}\ }\textbf {\bibinfo {volume} {34}},\
  \bibinfo {pages} {5390} (\bibinfo {year} {1986}{\natexlab{a}})}\BibitemShut
  {NoStop}%
\bibitem [{\citenamefont {Eiguren}\ \emph {et~al.}(2009)\citenamefont
  {Eiguren}, \citenamefont {Ambrosch-Draxl},\ and\ \citenamefont
  {Echenique}}]{eiguren_epc_2009}%
  \BibitemOpen
  \bibfield  {author} {\bibinfo {author} {\bibfnamefont {A.}~\bibnamefont
  {Eiguren}}, \bibinfo {author} {\bibfnamefont {C.}~\bibnamefont
  {Ambrosch-Draxl}}, \ and\ \bibinfo {author} {\bibfnamefont {P.~M.}\
  \bibnamefont {Echenique}},\ }\href {\doibase 10.1103/PhysRevB.79.245103}
  {\bibfield  {journal} {\bibinfo  {journal} {Phys. Rev. B}\ }\textbf {\bibinfo
  {volume} {79}},\ \bibinfo {pages} {245103} (\bibinfo {year}
  {2009})}\BibitemShut {NoStop}%
\bibitem [{\citenamefont {Giustino}(2017)}]{Giustino}%
  \BibitemOpen
  \bibfield  {author} {\bibinfo {author} {\bibfnamefont {F.}~\bibnamefont
  {Giustino}},\ }\href {\doibase 10.1103/RevModPhys.89.015003} {\bibfield
  {journal} {\bibinfo  {journal} {Rev. Mod. Phys.}\ }\textbf {\bibinfo {volume}
  {89}},\ \bibinfo {pages} {015003} (\bibinfo {year} {2017})}\BibitemShut
  {NoStop}%
\bibitem [{\citenamefont {Deslippe}\ \emph
  {et~al.}(2012{\natexlab{a}})\citenamefont {Deslippe}, \citenamefont
  {Samsonidze}, \citenamefont {Strubbe}, \citenamefont {Jain}, \citenamefont
  {Cohen},\ and\ \citenamefont {Louie}}]{DESLIPPE20121269}%
  \BibitemOpen
  \bibfield  {author} {\bibinfo {author} {\bibfnamefont {J.}~\bibnamefont
  {Deslippe}}, \bibinfo {author} {\bibfnamefont {G.}~\bibnamefont
  {Samsonidze}}, \bibinfo {author} {\bibfnamefont {D.~A.}\ \bibnamefont
  {Strubbe}}, \bibinfo {author} {\bibfnamefont {M.}~\bibnamefont {Jain}},
  \bibinfo {author} {\bibfnamefont {M.~L.}\ \bibnamefont {Cohen}}, \ and\
  \bibinfo {author} {\bibfnamefont {S.~G.}\ \bibnamefont {Louie}},\ }\href
  {\doibase https://doi.org/10.1016/j.cpc.2011.12.006} {\bibfield  {journal}
  {\bibinfo  {journal} {Computer Physics Communications}\ }\textbf {\bibinfo
  {volume} {183}},\ \bibinfo {pages} {1269} (\bibinfo {year}
  {2012}{\natexlab{a}})}\BibitemShut {NoStop}%
\bibitem [{\citenamefont {Strinati}(1988)}]{strinati1988application}%
  \BibitemOpen
  \bibfield  {author} {\bibinfo {author} {\bibfnamefont {G.}~\bibnamefont
  {Strinati}},\ }\href {\doibase 10.1007/BF02725962} {\bibfield  {journal}
  {\bibinfo  {journal} {Riv. Nuovo Cim.}\ }\textbf {\bibinfo {volume} {11}},\
  \bibinfo {pages} {1} (\bibinfo {year} {1988})}\BibitemShut {NoStop}%
\bibitem [{\citenamefont {Rohlfing}\ and\ \citenamefont
  {Louie}(2000)}]{rohlf_louie_2000}%
  \BibitemOpen
  \bibfield  {author} {\bibinfo {author} {\bibfnamefont {M.}~\bibnamefont
  {Rohlfing}}\ and\ \bibinfo {author} {\bibfnamefont {S.~G.}\ \bibnamefont
  {Louie}},\ }\href {\doibase 10.1103/PhysRevB.62.4927} {\bibfield  {journal}
  {\bibinfo  {journal} {Phys. Rev. B}\ }\textbf {\bibinfo {volume} {62}},\
  \bibinfo {pages} {4927} (\bibinfo {year} {2000})}\BibitemShut {NoStop}%
\bibitem [{\citenamefont {Rohlfing}\ and\ \citenamefont
  {Louie}(1998)}]{rohlf_louie_1998}%
  \BibitemOpen
  \bibfield  {author} {\bibinfo {author} {\bibfnamefont {M.}~\bibnamefont
  {Rohlfing}}\ and\ \bibinfo {author} {\bibfnamefont {S.~G.}\ \bibnamefont
  {Louie}},\ }\href {\doibase 10.1103/PhysRevLett.81.2312} {\bibfield
  {journal} {\bibinfo  {journal} {Phys. Rev. Lett.}\ }\textbf {\bibinfo
  {volume} {81}},\ \bibinfo {pages} {2312} (\bibinfo {year}
  {1998})}\BibitemShut {NoStop}%
\bibitem [{\citenamefont {Rudin}\ \emph {et~al.}(1990)\citenamefont {Rudin},
  \citenamefont {Reinecke},\ and\ \citenamefont
  {Segall}}]{rudin_exc_line_1990}%
  \BibitemOpen
  \bibfield  {author} {\bibinfo {author} {\bibfnamefont {S.}~\bibnamefont
  {Rudin}}, \bibinfo {author} {\bibfnamefont {T.~L.}\ \bibnamefont {Reinecke}},
  \ and\ \bibinfo {author} {\bibfnamefont {B.}~\bibnamefont {Segall}},\ }\href
  {\doibase 10.1103/PhysRevB.42.11218} {\bibfield  {journal} {\bibinfo
  {journal} {Phys. Rev. B}\ }\textbf {\bibinfo {volume} {42}},\ \bibinfo
  {pages} {11218} (\bibinfo {year} {1990})}\BibitemShut {NoStop}%
\bibitem [{\citenamefont {Chan}\ \emph
  {et~al.}(2023{\natexlab{a}})\citenamefont {Chan}, \citenamefont {Haber},
  \citenamefont {Naik}, \citenamefont {Neaton}, \citenamefont {Qiu},
  \citenamefont {da~Jornada},\ and\ \citenamefont {Louie}}]{Chan2023-jg}%
  \BibitemOpen
  \bibfield  {author} {\bibinfo {author} {\bibfnamefont {Y.-H.}\ \bibnamefont
  {Chan}}, \bibinfo {author} {\bibfnamefont {J.~B.}\ \bibnamefont {Haber}},
  \bibinfo {author} {\bibfnamefont {M.~H.}\ \bibnamefont {Naik}}, \bibinfo
  {author} {\bibfnamefont {J.~B.}\ \bibnamefont {Neaton}}, \bibinfo {author}
  {\bibfnamefont {D.~Y.}\ \bibnamefont {Qiu}}, \bibinfo {author} {\bibfnamefont
  {F.~H.}\ \bibnamefont {da~Jornada}}, \ and\ \bibinfo {author} {\bibfnamefont
  {S.~G.}\ \bibnamefont {Louie}},\ }\href@noop {} {\bibfield  {journal}
  {\bibinfo  {journal} {Nano Lett.}\ }\textbf {\bibinfo {volume} {23}},\
  \bibinfo {pages} {3971} (\bibinfo {year} {2023}{\natexlab{a}})}\BibitemShut
  {NoStop}%
\bibitem [{\citenamefont {Marini}(2008)}]{marini_finite_t_exc_2008}%
  \BibitemOpen
  \bibfield  {author} {\bibinfo {author} {\bibfnamefont {A.}~\bibnamefont
  {Marini}},\ }\href {\doibase 10.1103/PhysRevLett.101.106405} {\bibfield
  {journal} {\bibinfo  {journal} {Phys. Rev. Lett.}\ }\textbf {\bibinfo
  {volume} {101}},\ \bibinfo {pages} {106405} (\bibinfo {year}
  {2008})}\BibitemShut {NoStop}%
\bibitem [{\citenamefont {D’innocenzo}\ \emph {et~al.}(2014)\citenamefont
  {D’innocenzo}, \citenamefont {Grancini}, \citenamefont {Alcocer},
  \citenamefont {Kandada}, \citenamefont {Stranks}, \citenamefont {Lee},
  \citenamefont {Lanzani}, \citenamefont {Snaith},\ and\ \citenamefont
  {Petrozza}}]{exciton_perovskites}%
  \BibitemOpen
  \bibfield  {author} {\bibinfo {author} {\bibfnamefont {V.}~\bibnamefont
  {D’innocenzo}}, \bibinfo {author} {\bibfnamefont {G.}~\bibnamefont
  {Grancini}}, \bibinfo {author} {\bibfnamefont {M.~J.}\ \bibnamefont
  {Alcocer}}, \bibinfo {author} {\bibfnamefont {A.~R.~S.}\ \bibnamefont
  {Kandada}}, \bibinfo {author} {\bibfnamefont {S.~D.}\ \bibnamefont
  {Stranks}}, \bibinfo {author} {\bibfnamefont {M.~M.}\ \bibnamefont {Lee}},
  \bibinfo {author} {\bibfnamefont {G.}~\bibnamefont {Lanzani}}, \bibinfo
  {author} {\bibfnamefont {H.~J.}\ \bibnamefont {Snaith}}, \ and\ \bibinfo
  {author} {\bibfnamefont {A.}~\bibnamefont {Petrozza}},\ }\href {\doibase
  10.1038/ncomms4586} {\bibfield  {journal} {\bibinfo  {journal} {Nature
  communications}\ }\textbf {\bibinfo {volume} {5}},\ \bibinfo {pages} {3586}
  (\bibinfo {year} {2014})}\BibitemShut {NoStop}%
\bibitem [{\citenamefont {Iadonisi}\ and\ \citenamefont
  {Bassani}(1987)}]{Iadonisi1987}%
  \BibitemOpen
  \bibfield  {author} {\bibinfo {author} {\bibfnamefont {G.}~\bibnamefont
  {Iadonisi}}\ and\ \bibinfo {author} {\bibfnamefont {F.}~\bibnamefont
  {Bassani}},\ }\href {\doibase 10.1007/bf02457030} {\bibfield  {journal}
  {\bibinfo  {journal} {Il Nuovo Cimento D}\ }\textbf {\bibinfo {volume} {9}},\
  \bibinfo {pages} {703–714} (\bibinfo {year} {1987})}\BibitemShut {NoStop}%
\bibitem [{\citenamefont {Dai}\ \emph {et~al.}(2024{\natexlab{a}})\citenamefont
  {Dai}, \citenamefont {Lian}, \citenamefont {Lafuente-Bartolome},\ and\
  \citenamefont {Giustino}}]{giustino_exc_polarons_24_prl}%
  \BibitemOpen
  \bibfield  {author} {\bibinfo {author} {\bibfnamefont {Z.}~\bibnamefont
  {Dai}}, \bibinfo {author} {\bibfnamefont {C.}~\bibnamefont {Lian}}, \bibinfo
  {author} {\bibfnamefont {J.}~\bibnamefont {Lafuente-Bartolome}}, \ and\
  \bibinfo {author} {\bibfnamefont {F.}~\bibnamefont {Giustino}},\ }\href
  {\doibase 10.1103/PhysRevLett.132.036902} {\bibfield  {journal} {\bibinfo
  {journal} {Phys. Rev. Lett.}\ }\textbf {\bibinfo {volume} {132}},\ \bibinfo
  {pages} {036902} (\bibinfo {year} {2024}{\natexlab{a}})}\BibitemShut
  {NoStop}%
\bibitem [{\citenamefont {Noffsinger}\ \emph {et~al.}(2012)\citenamefont
  {Noffsinger}, \citenamefont {Kioupakis}, \citenamefont {Van~de Walle},
  \citenamefont {Louie},\ and\ \citenamefont {Cohen}}]{noffsinger_opt_ph_2012}%
  \BibitemOpen
  \bibfield  {author} {\bibinfo {author} {\bibfnamefont {J.}~\bibnamefont
  {Noffsinger}}, \bibinfo {author} {\bibfnamefont {E.}~\bibnamefont
  {Kioupakis}}, \bibinfo {author} {\bibfnamefont {C.~G.}\ \bibnamefont {Van~de
  Walle}}, \bibinfo {author} {\bibfnamefont {S.~G.}\ \bibnamefont {Louie}}, \
  and\ \bibinfo {author} {\bibfnamefont {M.~L.}\ \bibnamefont {Cohen}},\ }\href
  {\doibase 10.1103/PhysRevLett.108.167402} {\bibfield  {journal} {\bibinfo
  {journal} {Phys. Rev. Lett.}\ }\textbf {\bibinfo {volume} {108}},\ \bibinfo
  {pages} {167402} (\bibinfo {year} {2012})}\BibitemShut {NoStop}%
\bibitem [{\citenamefont {Zacharias}\ \emph {et~al.}(2015)\citenamefont
  {Zacharias}, \citenamefont {Patrick},\ and\ \citenamefont
  {Giustino}}]{zacharias_ph_opt_2015}%
  \BibitemOpen
  \bibfield  {author} {\bibinfo {author} {\bibfnamefont {M.}~\bibnamefont
  {Zacharias}}, \bibinfo {author} {\bibfnamefont {C.~E.}\ \bibnamefont
  {Patrick}}, \ and\ \bibinfo {author} {\bibfnamefont {F.}~\bibnamefont
  {Giustino}},\ }\href {\doibase 10.1103/PhysRevLett.115.177401} {\bibfield
  {journal} {\bibinfo  {journal} {Phys. Rev. Lett.}\ }\textbf {\bibinfo
  {volume} {115}},\ \bibinfo {pages} {177401} (\bibinfo {year}
  {2015})}\BibitemShut {NoStop}%
\bibitem [{\citenamefont {Patrick}\ and\ \citenamefont
  {Giustino}(2014)}]{Patrick2014}%
  \BibitemOpen
  \bibfield  {author} {\bibinfo {author} {\bibfnamefont {C.~E.}\ \bibnamefont
  {Patrick}}\ and\ \bibinfo {author} {\bibfnamefont {F.}~\bibnamefont
  {Giustino}},\ }\href {\doibase 10.1088/0953-8984/26/36/365503} {\bibfield
  {journal} {\bibinfo  {journal} {Journal of Physics: Condensed Matter}\
  }\textbf {\bibinfo {volume} {26}},\ \bibinfo {pages} {365503} (\bibinfo
  {year} {2014})}\BibitemShut {NoStop}%
\bibitem [{\citenamefont {Helmrich}\ \emph {et~al.}(2021)\citenamefont
  {Helmrich}, \citenamefont {Sampson}, \citenamefont {Huang}, \citenamefont
  {Selig}, \citenamefont {Hao}, \citenamefont {Tran}, \citenamefont {Achstein},
  \citenamefont {Young}, \citenamefont {Knorr}, \citenamefont {Malic},
  \citenamefont {Woggon}, \citenamefont {Owschimikow},\ and\ \citenamefont
  {Li}}]{helmrich_2021}%
  \BibitemOpen
  \bibfield  {author} {\bibinfo {author} {\bibfnamefont {S.}~\bibnamefont
  {Helmrich}}, \bibinfo {author} {\bibfnamefont {K.}~\bibnamefont {Sampson}},
  \bibinfo {author} {\bibfnamefont {D.}~\bibnamefont {Huang}}, \bibinfo
  {author} {\bibfnamefont {M.}~\bibnamefont {Selig}}, \bibinfo {author}
  {\bibfnamefont {K.}~\bibnamefont {Hao}}, \bibinfo {author} {\bibfnamefont
  {K.}~\bibnamefont {Tran}}, \bibinfo {author} {\bibfnamefont {A.}~\bibnamefont
  {Achstein}}, \bibinfo {author} {\bibfnamefont {C.}~\bibnamefont {Young}},
  \bibinfo {author} {\bibfnamefont {A.}~\bibnamefont {Knorr}}, \bibinfo
  {author} {\bibfnamefont {E.}~\bibnamefont {Malic}}, \bibinfo {author}
  {\bibfnamefont {U.}~\bibnamefont {Woggon}}, \bibinfo {author} {\bibfnamefont
  {N.}~\bibnamefont {Owschimikow}}, \ and\ \bibinfo {author} {\bibfnamefont
  {X.}~\bibnamefont {Li}},\ }\href {\doibase 10.1103/PhysRevLett.127.157403}
  {\bibfield  {journal} {\bibinfo  {journal} {Phys. Rev. Lett.}\ }\textbf
  {\bibinfo {volume} {127}},\ \bibinfo {pages} {157403} (\bibinfo {year}
  {2021})}\BibitemShut {NoStop}%
\bibitem [{\citenamefont {Bechstedt}(2016)}]{bechstedt2016many}%
  \BibitemOpen
  \bibfield  {author} {\bibinfo {author} {\bibfnamefont {F.}~\bibnamefont
  {Bechstedt}},\ }\href {\doibase 10.1007/978-3-662-44593-8} {\emph {\bibinfo
  {title} {Many-Body Approach to Electronic Excitations}}}\ (\bibinfo
  {publisher} {Springer},\ \bibinfo {address} {Berlin},\ \bibinfo {year}
  {2016})\BibitemShut {NoStop}%
\bibitem [{\citenamefont {Shree}\ \emph {et~al.}(2018)\citenamefont {Shree},
  \citenamefont {Semina}, \citenamefont {Robert}, \citenamefont {Han},
  \citenamefont {Amand}, \citenamefont {Balocchi}, \citenamefont {Manca},
  \citenamefont {Courtade}, \citenamefont {Marie}, \citenamefont {Taniguchi},
  \citenamefont {Watanabe}, \citenamefont {Glazov},\ and\ \citenamefont
  {Urbaszek}}]{shree_excph_mose2_2018}%
  \BibitemOpen
  \bibfield  {author} {\bibinfo {author} {\bibfnamefont {S.}~\bibnamefont
  {Shree}}, \bibinfo {author} {\bibfnamefont {M.}~\bibnamefont {Semina}},
  \bibinfo {author} {\bibfnamefont {C.}~\bibnamefont {Robert}}, \bibinfo
  {author} {\bibfnamefont {B.}~\bibnamefont {Han}}, \bibinfo {author}
  {\bibfnamefont {T.}~\bibnamefont {Amand}}, \bibinfo {author} {\bibfnamefont
  {A.}~\bibnamefont {Balocchi}}, \bibinfo {author} {\bibfnamefont
  {M.}~\bibnamefont {Manca}}, \bibinfo {author} {\bibfnamefont
  {E.}~\bibnamefont {Courtade}}, \bibinfo {author} {\bibfnamefont
  {X.}~\bibnamefont {Marie}}, \bibinfo {author} {\bibfnamefont
  {T.}~\bibnamefont {Taniguchi}}, \bibinfo {author} {\bibfnamefont
  {K.}~\bibnamefont {Watanabe}}, \bibinfo {author} {\bibfnamefont {M.~M.}\
  \bibnamefont {Glazov}}, \ and\ \bibinfo {author} {\bibfnamefont
  {B.}~\bibnamefont {Urbaszek}},\ }\href {\doibase 10.1103/PhysRevB.98.035302}
  {\bibfield  {journal} {\bibinfo  {journal} {Phys. Rev. B}\ }\textbf {\bibinfo
  {volume} {98}},\ \bibinfo {pages} {035302} (\bibinfo {year}
  {2018})}\BibitemShut {NoStop}%
\bibitem [{\citenamefont {Tassone}\ and\ \citenamefont
  {Yamamoto}(1999)}]{tassone_1999}%
  \BibitemOpen
  \bibfield  {author} {\bibinfo {author} {\bibfnamefont {F.}~\bibnamefont
  {Tassone}}\ and\ \bibinfo {author} {\bibfnamefont {Y.}~\bibnamefont
  {Yamamoto}},\ }\href {\doibase 10.1103/PhysRevB.59.10830} {\bibfield
  {journal} {\bibinfo  {journal} {Phys. Rev. B}\ }\textbf {\bibinfo {volume}
  {59}},\ \bibinfo {pages} {10830} (\bibinfo {year} {1999})}\BibitemShut
  {NoStop}%
\bibitem [{\citenamefont {Iadonisi}\ and\ \citenamefont
  {Bassani}(1983)}]{Iadonisi1983}%
  \BibitemOpen
  \bibfield  {author} {\bibinfo {author} {\bibfnamefont {G.}~\bibnamefont
  {Iadonisi}}\ and\ \bibinfo {author} {\bibfnamefont {F.}~\bibnamefont
  {Bassani}},\ }\href {\doibase 10.1007/bf02460231} {\bibfield  {journal}
  {\bibinfo  {journal} {Il Nuovo Cimento D}\ }\textbf {\bibinfo {volume} {2}},\
  \bibinfo {pages} {1541–1560} (\bibinfo {year} {1983})}\BibitemShut
  {NoStop}%
\bibitem [{\citenamefont {Toyozawa}(1958)}]{Toyozawa_58}%
  \BibitemOpen
  \bibfield  {author} {\bibinfo {author} {\bibfnamefont {Y.}~\bibnamefont
  {Toyozawa}},\ }\href {\doibase 10.1143/PTP.20.53} {\bibfield  {journal}
  {\bibinfo  {journal} {Progress of Theoretical Physics}\ }\textbf {\bibinfo
  {volume} {20}},\ \bibinfo {pages} {53} (\bibinfo {year} {1958})},\ \Eprint
  {http://arxiv.org/abs/https://academic.oup.com/ptp/article-pdf/20/1/53/5457877/20-1-53.pdf}
  {https://academic.oup.com/ptp/article-pdf/20/1/53/5457877/20-1-53.pdf}
  \BibitemShut {NoStop}%
\bibitem [{\citenamefont {Yarkony}\ and\ \citenamefont
  {Silbey}(1976)}]{yarkony_76}%
  \BibitemOpen
  \bibfield  {author} {\bibinfo {author} {\bibfnamefont {D.}~\bibnamefont
  {Yarkony}}\ and\ \bibinfo {author} {\bibfnamefont {R.}~\bibnamefont
  {Silbey}},\ }\href {\doibase 10.1063/1.433182} {\bibfield  {journal}
  {\bibinfo  {journal} {The Journal of Chemical Physics}\ }\textbf {\bibinfo
  {volume} {65}},\ \bibinfo {pages} {1042} (\bibinfo {year} {1976})},\ \Eprint
  {http://arxiv.org/abs/https://pubs.aip.org/aip/jcp/article-pdf/65/3/1042/18901579/1042\_1\_online.pdf}
  {https://pubs.aip.org/aip/jcp/article-pdf/65/3/1042/18901579/1042\_1\_online.pdf}
  \BibitemShut {NoStop}%
\bibitem [{\citenamefont {Pollmann}\ and\ \citenamefont
  {B\"uttner}(1977)}]{pollmann_1977}%
  \BibitemOpen
  \bibfield  {author} {\bibinfo {author} {\bibfnamefont {J.}~\bibnamefont
  {Pollmann}}\ and\ \bibinfo {author} {\bibfnamefont {H.}~\bibnamefont
  {B\"uttner}},\ }\href {\doibase 10.1103/PhysRevB.16.4480} {\bibfield
  {journal} {\bibinfo  {journal} {Phys. Rev. B}\ }\textbf {\bibinfo {volume}
  {16}},\ \bibinfo {pages} {4480} (\bibinfo {year} {1977})}\BibitemShut
  {NoStop}%
\bibitem [{\citenamefont {Mahanti}\ and\ \citenamefont
  {Varma}(1970)}]{mahanti_eff_el_1970}%
  \BibitemOpen
  \bibfield  {author} {\bibinfo {author} {\bibfnamefont {S.~D.}\ \bibnamefont
  {Mahanti}}\ and\ \bibinfo {author} {\bibfnamefont {C.~M.}\ \bibnamefont
  {Varma}},\ }\href {\doibase 10.1103/PhysRevLett.25.1115} {\bibfield
  {journal} {\bibinfo  {journal} {Phys. Rev. Lett.}\ }\textbf {\bibinfo
  {volume} {25}},\ \bibinfo {pages} {1115} (\bibinfo {year}
  {1970})}\BibitemShut {NoStop}%
\bibitem [{\citenamefont {Mahanti}\ and\ \citenamefont
  {Varma}(1972)}]{mahanti_eff_el_1972}%
  \BibitemOpen
  \bibfield  {author} {\bibinfo {author} {\bibfnamefont {S.~D.}\ \bibnamefont
  {Mahanti}}\ and\ \bibinfo {author} {\bibfnamefont {C.~M.}\ \bibnamefont
  {Varma}},\ }\href {\doibase 10.1103/PhysRevB.6.2209} {\bibfield  {journal}
  {\bibinfo  {journal} {Phys. Rev. B}\ }\textbf {\bibinfo {volume} {6}},\
  \bibinfo {pages} {2209} (\bibinfo {year} {1972})}\BibitemShut {NoStop}%
\bibitem [{\citenamefont {Verzelen}\ \emph {et~al.}(2002)\citenamefont
  {Verzelen}, \citenamefont {Ferreira},\ and\ \citenamefont
  {Bastard}}]{verzelen_exc_pol_2022}%
  \BibitemOpen
  \bibfield  {author} {\bibinfo {author} {\bibfnamefont {O.}~\bibnamefont
  {Verzelen}}, \bibinfo {author} {\bibfnamefont {R.}~\bibnamefont {Ferreira}},
  \ and\ \bibinfo {author} {\bibfnamefont {G.}~\bibnamefont {Bastard}},\ }\href
  {\doibase 10.1103/PhysRevLett.88.146803} {\bibfield  {journal} {\bibinfo
  {journal} {Phys. Rev. Lett.}\ }\textbf {\bibinfo {volume} {88}},\ \bibinfo
  {pages} {146803} (\bibinfo {year} {2002})}\BibitemShut {NoStop}%
\bibitem [{\citenamefont {Bechstedt}\ and\ \citenamefont
  {Furthmüller}(2019)}]{bechstedt_ga203}%
  \BibitemOpen
  \bibfield  {author} {\bibinfo {author} {\bibfnamefont {F.}~\bibnamefont
  {Bechstedt}}\ and\ \bibinfo {author} {\bibfnamefont {J.}~\bibnamefont
  {Furthmüller}},\ }\href {\doibase 10.1063/1.5084324} {\bibfield  {journal}
  {\bibinfo  {journal} {Applied Physics Letters}\ }\textbf {\bibinfo {volume}
  {114}},\ \bibinfo {pages} {122101} (\bibinfo {year} {2019})}\BibitemShut
  {NoStop}%
\bibitem [{\citenamefont {Schleife}\ \emph {et~al.}(2018)\citenamefont
  {Schleife}, \citenamefont {Neumann}, \citenamefont {Esser}, \citenamefont
  {Galazka}, \citenamefont {Gottwald}, \citenamefont {Nixdorf}, \citenamefont
  {Goldhahn},\ and\ \citenamefont {Feneberg}}]{Schleife2018}%
  \BibitemOpen
  \bibfield  {author} {\bibinfo {author} {\bibfnamefont {A.}~\bibnamefont
  {Schleife}}, \bibinfo {author} {\bibfnamefont {M.~D.}\ \bibnamefont
  {Neumann}}, \bibinfo {author} {\bibfnamefont {N.}~\bibnamefont {Esser}},
  \bibinfo {author} {\bibfnamefont {Z.}~\bibnamefont {Galazka}}, \bibinfo
  {author} {\bibfnamefont {A.}~\bibnamefont {Gottwald}}, \bibinfo {author}
  {\bibfnamefont {J.}~\bibnamefont {Nixdorf}}, \bibinfo {author} {\bibfnamefont
  {R.}~\bibnamefont {Goldhahn}}, \ and\ \bibinfo {author} {\bibfnamefont
  {M.}~\bibnamefont {Feneberg}},\ }\href {\doibase 10.1088/1367-2630/aabeb0}
  {\bibfield  {journal} {\bibinfo  {journal} {New Journal of Physics}\ }\textbf
  {\bibinfo {volume} {20}},\ \bibinfo {pages} {053016} (\bibinfo {year}
  {2018})}\BibitemShut {NoStop}%
\bibitem [{\citenamefont {Park}\ and\ \citenamefont {Limmer}(2022)}]{Park2022}%
  \BibitemOpen
  \bibfield  {author} {\bibinfo {author} {\bibfnamefont {Y.}~\bibnamefont
  {Park}}\ and\ \bibinfo {author} {\bibfnamefont {D.~T.}\ \bibnamefont
  {Limmer}},\ }\href {\doibase 10.1063/5.0100738} {\bibfield  {journal}
  {\bibinfo  {journal} {The Journal of Chemical Physics}\ }\textbf {\bibinfo
  {volume} {157}} (\bibinfo {year} {2022}),\ 10.1063/5.0100738}\BibitemShut
  {NoStop}%
\bibitem [{\citenamefont {Yang}\ and\ \citenamefont
  {Draxl}(2022)}]{yang_exc_struc_relax_22}%
  \BibitemOpen
  \bibfield  {author} {\bibinfo {author} {\bibfnamefont {M.}~\bibnamefont
  {Yang}}\ and\ \bibinfo {author} {\bibfnamefont {C.}~\bibnamefont {Draxl}},\
  }\href {\doibase 10.48550/ARXIV.2212.13645} {\enquote {\bibinfo {title}
  {Novel approach to structural relaxation of materials in optically excited
  states},}\ } (\bibinfo {year} {2022})\BibitemShut {NoStop}%
\bibitem [{\citenamefont {Chen}\ \emph {et~al.}(2020)\citenamefont {Chen},
  \citenamefont {Sangalli},\ and\ \citenamefont {Bernardi}}]{chen_exc_ph_20}%
  \BibitemOpen
  \bibfield  {author} {\bibinfo {author} {\bibfnamefont {H.-Y.}\ \bibnamefont
  {Chen}}, \bibinfo {author} {\bibfnamefont {D.}~\bibnamefont {Sangalli}}, \
  and\ \bibinfo {author} {\bibfnamefont {M.}~\bibnamefont {Bernardi}},\ }\href
  {\doibase 10.1103/PhysRevLett.125.107401} {\bibfield  {journal} {\bibinfo
  {journal} {Phys. Rev. Lett.}\ }\textbf {\bibinfo {volume} {125}},\ \bibinfo
  {pages} {107401} (\bibinfo {year} {2020})}\BibitemShut {NoStop}%
\bibitem [{\citenamefont {Cudazzo}(2020)}]{cudazzo_2020}%
  \BibitemOpen
  \bibfield  {author} {\bibinfo {author} {\bibfnamefont {P.}~\bibnamefont
  {Cudazzo}},\ }\href {\doibase 10.1103/PhysRevB.102.045136} {\bibfield
  {journal} {\bibinfo  {journal} {Phys. Rev. B}\ }\textbf {\bibinfo {volume}
  {102}},\ \bibinfo {pages} {045136} (\bibinfo {year} {2020})}\BibitemShut
  {NoStop}%
\bibitem [{\citenamefont {Antonius}\ and\ \citenamefont
  {Louie}(2022)}]{antonius_exc_ph}%
  \BibitemOpen
  \bibfield  {author} {\bibinfo {author} {\bibfnamefont {G.}~\bibnamefont
  {Antonius}}\ and\ \bibinfo {author} {\bibfnamefont {S.~G.}\ \bibnamefont
  {Louie}},\ }\href {\doibase 10.1103/PhysRevB.105.085111} {\bibfield
  {journal} {\bibinfo  {journal} {Phys. Rev. B}\ }\textbf {\bibinfo {volume}
  {105}},\ \bibinfo {pages} {085111} (\bibinfo {year} {2022})}\BibitemShut
  {NoStop}%
\bibitem [{\citenamefont {Paleari}\ and\ \citenamefont
  {Marini}(2022)}]{paleari_exc_ph_22}%
  \BibitemOpen
  \bibfield  {author} {\bibinfo {author} {\bibfnamefont {F.}~\bibnamefont
  {Paleari}}\ and\ \bibinfo {author} {\bibfnamefont {A.}~\bibnamefont
  {Marini}},\ }\href {\doibase 10.1103/PhysRevB.106.125403} {\bibfield
  {journal} {\bibinfo  {journal} {Phys. Rev. B}\ }\textbf {\bibinfo {volume}
  {106}},\ \bibinfo {pages} {125403} (\bibinfo {year} {2022})}\BibitemShut
  {NoStop}%
\bibitem [{\citenamefont {Chan}\ \emph
  {et~al.}(2023{\natexlab{b}})\citenamefont {Chan}, \citenamefont {Haber},
  \citenamefont {Naik}, \citenamefont {Neaton}, \citenamefont {Qiu},
  \citenamefont {da~Jornada},\ and\ \citenamefont
  {Louie}}]{Chan2023_exc_life_line}%
  \BibitemOpen
  \bibfield  {author} {\bibinfo {author} {\bibfnamefont {Y.-h.}\ \bibnamefont
  {Chan}}, \bibinfo {author} {\bibfnamefont {J.~B.}\ \bibnamefont {Haber}},
  \bibinfo {author} {\bibfnamefont {M.~H.}\ \bibnamefont {Naik}}, \bibinfo
  {author} {\bibfnamefont {J.~B.}\ \bibnamefont {Neaton}}, \bibinfo {author}
  {\bibfnamefont {D.~Y.}\ \bibnamefont {Qiu}}, \bibinfo {author} {\bibfnamefont
  {F.~H.}\ \bibnamefont {da~Jornada}}, \ and\ \bibinfo {author} {\bibfnamefont
  {S.~G.}\ \bibnamefont {Louie}},\ }\href {\doibase
  10.1021/acs.nanolett.3c00732} {\bibfield  {journal} {\bibinfo  {journal}
  {Nano Letters}\ }\textbf {\bibinfo {volume} {23}},\ \bibinfo {pages}
  {3971–3977} (\bibinfo {year} {2023}{\natexlab{b}})}\BibitemShut {NoStop}%
\bibitem [{\citenamefont {Filip}\ \emph {et~al.}(2021)\citenamefont {Filip},
  \citenamefont {Haber},\ and\ \citenamefont {Neaton}}]{filip_perovksites}%
  \BibitemOpen
  \bibfield  {author} {\bibinfo {author} {\bibfnamefont {M.~R.}\ \bibnamefont
  {Filip}}, \bibinfo {author} {\bibfnamefont {J.~B.}\ \bibnamefont {Haber}}, \
  and\ \bibinfo {author} {\bibfnamefont {J.~B.}\ \bibnamefont {Neaton}},\
  }\href {\doibase 10.1103/PhysRevLett.127.067401} {\bibfield  {journal}
  {\bibinfo  {journal} {Phys. Rev. Lett.}\ }\textbf {\bibinfo {volume} {127}},\
  \bibinfo {pages} {067401} (\bibinfo {year} {2021})}\BibitemShut {NoStop}%
\bibitem [{\citenamefont {Alvertis}\ \emph
  {et~al.}(2023{\natexlab{a}})\citenamefont {Alvertis}, \citenamefont {Haber},
  \citenamefont {Li}, \citenamefont {Coveney}, \citenamefont {Louie},
  \citenamefont {Filip},\ and\ \citenamefont {Neaton}}]{alvertis_bse_exc_2023}%
  \BibitemOpen
  \bibfield  {author} {\bibinfo {author} {\bibfnamefont {A.~M.}\ \bibnamefont
  {Alvertis}}, \bibinfo {author} {\bibfnamefont {J.~B.}\ \bibnamefont {Haber}},
  \bibinfo {author} {\bibfnamefont {Z.}~\bibnamefont {Li}}, \bibinfo {author}
  {\bibfnamefont {C.~J.~N.}\ \bibnamefont {Coveney}}, \bibinfo {author}
  {\bibfnamefont {S.~G.}\ \bibnamefont {Louie}}, \bibinfo {author}
  {\bibfnamefont {M.~R.}\ \bibnamefont {Filip}}, \ and\ \bibinfo {author}
  {\bibfnamefont {J.~B.}\ \bibnamefont {Neaton}},\ }\href@noop {} {\enquote
  {\bibinfo {title} {Phonon screening and dissociation of excitons at finite
  temperatures from first principles},}\ } (\bibinfo {year}
  {2023}{\natexlab{a}}),\ \Eprint {http://arxiv.org/abs/arXiv:2312.03841}
  {arXiv:2312.03841} \BibitemShut {NoStop}%
\bibitem [{\citenamefont {Gillet}\ \emph {et~al.}(2017)\citenamefont {Gillet},
  \citenamefont {Kontur}, \citenamefont {Giantomassi}, \citenamefont {Draxl},\
  and\ \citenamefont {Gonze}}]{Gillet2017}%
  \BibitemOpen
  \bibfield  {author} {\bibinfo {author} {\bibfnamefont {Y.}~\bibnamefont
  {Gillet}}, \bibinfo {author} {\bibfnamefont {S.}~\bibnamefont {Kontur}},
  \bibinfo {author} {\bibfnamefont {M.}~\bibnamefont {Giantomassi}}, \bibinfo
  {author} {\bibfnamefont {C.}~\bibnamefont {Draxl}}, \ and\ \bibinfo {author}
  {\bibfnamefont {X.}~\bibnamefont {Gonze}},\ }\href {\doibase
  10.1038/s41598-017-07682-y} {\bibfield  {journal} {\bibinfo  {journal}
  {Scientific Reports}\ }\textbf {\bibinfo {volume} {7}} (\bibinfo {year}
  {2017}),\ 10.1038/s41598-017-07682-y}\BibitemShut {NoStop}%
\bibitem [{\citenamefont {Olovsson}\ \emph {et~al.}(2019)\citenamefont
  {Olovsson}, \citenamefont {Mizoguchi}, \citenamefont {Magnuson},
  \citenamefont {Kontur}, \citenamefont {Hellman}, \citenamefont {Tanaka},\
  and\ \citenamefont {Draxl}}]{Olovsson2019}%
  \BibitemOpen
  \bibfield  {author} {\bibinfo {author} {\bibfnamefont {W.}~\bibnamefont
  {Olovsson}}, \bibinfo {author} {\bibfnamefont {T.}~\bibnamefont {Mizoguchi}},
  \bibinfo {author} {\bibfnamefont {M.}~\bibnamefont {Magnuson}}, \bibinfo
  {author} {\bibfnamefont {S.}~\bibnamefont {Kontur}}, \bibinfo {author}
  {\bibfnamefont {O.}~\bibnamefont {Hellman}}, \bibinfo {author} {\bibfnamefont
  {I.}~\bibnamefont {Tanaka}}, \ and\ \bibinfo {author} {\bibfnamefont
  {C.}~\bibnamefont {Draxl}},\ }\href {\doibase 10.1021/acs.jpcc.9b00179}
  {\bibfield  {journal} {\bibinfo  {journal} {The Journal of Physical Chemistry
  C}\ }\textbf {\bibinfo {volume} {123}},\ \bibinfo {pages} {9688–9692}
  (\bibinfo {year} {2019})}\BibitemShut {NoStop}%
\bibitem [{\citenamefont {Giannozzi}\ \emph {et~al.}(1991)\citenamefont
  {Giannozzi}, \citenamefont {de~Gironcoli}, \citenamefont {Pavone},\ and\
  \citenamefont {Baroni}}]{pavone}%
  \BibitemOpen
  \bibfield  {author} {\bibinfo {author} {\bibfnamefont {P.}~\bibnamefont
  {Giannozzi}}, \bibinfo {author} {\bibfnamefont {S.}~\bibnamefont
  {de~Gironcoli}}, \bibinfo {author} {\bibfnamefont {P.}~\bibnamefont
  {Pavone}}, \ and\ \bibinfo {author} {\bibfnamefont {S.}~\bibnamefont
  {Baroni}},\ }\href {\doibase 10.1103/PhysRevB.43.7231} {\bibfield  {journal}
  {\bibinfo  {journal} {Phys. Rev. B}\ }\textbf {\bibinfo {volume} {43}},\
  \bibinfo {pages} {7231} (\bibinfo {year} {1991})}\BibitemShut {NoStop}%
\bibitem [{\citenamefont {Fuchs}\ \emph {et~al.}(2008)\citenamefont {Fuchs},
  \citenamefont {R\"odl}, \citenamefont {Schleife},\ and\ \citenamefont
  {Bechstedt}}]{fuchs_08}%
  \BibitemOpen
  \bibfield  {author} {\bibinfo {author} {\bibfnamefont {F.}~\bibnamefont
  {Fuchs}}, \bibinfo {author} {\bibfnamefont {C.}~\bibnamefont {R\"odl}},
  \bibinfo {author} {\bibfnamefont {A.}~\bibnamefont {Schleife}}, \ and\
  \bibinfo {author} {\bibfnamefont {F.}~\bibnamefont {Bechstedt}},\ }\href
  {\doibase 10.1103/PhysRevB.78.085103} {\bibfield  {journal} {\bibinfo
  {journal} {Phys. Rev. B}\ }\textbf {\bibinfo {volume} {78}},\ \bibinfo
  {pages} {085103} (\bibinfo {year} {2008})}\BibitemShut {NoStop}%
\bibitem [{\citenamefont {Adamska}\ and\ \citenamefont {Umari}(2021)}]{umari}%
  \BibitemOpen
  \bibfield  {author} {\bibinfo {author} {\bibfnamefont {L.}~\bibnamefont
  {Adamska}}\ and\ \bibinfo {author} {\bibfnamefont {P.}~\bibnamefont
  {Umari}},\ }\href {\doibase 10.1103/PhysRevB.103.075201} {\bibfield
  {journal} {\bibinfo  {journal} {Phys. Rev. B}\ }\textbf {\bibinfo {volume}
  {103}},\ \bibinfo {pages} {075201} (\bibinfo {year} {2021})}\BibitemShut
  {NoStop}%
\bibitem [{\citenamefont {Umari}\ \emph {et~al.}(2018)\citenamefont {Umari},
  \citenamefont {Mosconi},\ and\ \citenamefont
  {De~Angelis}}]{umari_perovskites}%
  \BibitemOpen
  \bibfield  {author} {\bibinfo {author} {\bibfnamefont {P.}~\bibnamefont
  {Umari}}, \bibinfo {author} {\bibfnamefont {E.}~\bibnamefont {Mosconi}}, \
  and\ \bibinfo {author} {\bibfnamefont {F.}~\bibnamefont {De~Angelis}},\
  }\href {\doibase 10.1021/acs.jpclett.7b03286} {\bibfield  {journal} {\bibinfo
   {journal} {The Journal of Physical Chemistry Letters}\ }\textbf {\bibinfo
  {volume} {9}},\ \bibinfo {pages} {620} (\bibinfo {year} {2018})},\ \bibinfo
  {note} {pMID: 29336156},\ \Eprint
  {http://arxiv.org/abs/https://doi.org/10.1021/acs.jpclett.7b03286}
  {https://doi.org/10.1021/acs.jpclett.7b03286} \BibitemShut {NoStop}%
\bibitem [{\citenamefont {Bechstedt}\ \emph {et~al.}(2005)\citenamefont
  {Bechstedt}, \citenamefont {Seino}, \citenamefont {Hahn},\ and\ \citenamefont
  {Schmidt}}]{highly_ionic}%
  \BibitemOpen
  \bibfield  {author} {\bibinfo {author} {\bibfnamefont {F.}~\bibnamefont
  {Bechstedt}}, \bibinfo {author} {\bibfnamefont {K.}~\bibnamefont {Seino}},
  \bibinfo {author} {\bibfnamefont {P.~H.}\ \bibnamefont {Hahn}}, \ and\
  \bibinfo {author} {\bibfnamefont {W.~G.}\ \bibnamefont {Schmidt}},\ }\href
  {\doibase 10.1103/PhysRevB.72.245114} {\bibfield  {journal} {\bibinfo
  {journal} {Phys. Rev. B}\ }\textbf {\bibinfo {volume} {72}},\ \bibinfo
  {pages} {245114} (\bibinfo {year} {2005})}\BibitemShut {NoStop}%
\bibitem [{\citenamefont {Bokdam}\ \emph {et~al.}(2016)\citenamefont {Bokdam},
  \citenamefont {Sander}, \citenamefont {Stroppa}, \citenamefont {Picozzi},
  \citenamefont {Sarma}, \citenamefont {Franchini},\ and\ \citenamefont
  {Kresse}}]{bokdam2016role}%
  \BibitemOpen
  \bibfield  {author} {\bibinfo {author} {\bibfnamefont {M.}~\bibnamefont
  {Bokdam}}, \bibinfo {author} {\bibfnamefont {T.}~\bibnamefont {Sander}},
  \bibinfo {author} {\bibfnamefont {A.}~\bibnamefont {Stroppa}}, \bibinfo
  {author} {\bibfnamefont {S.}~\bibnamefont {Picozzi}}, \bibinfo {author}
  {\bibfnamefont {D.}~\bibnamefont {Sarma}}, \bibinfo {author} {\bibfnamefont
  {C.}~\bibnamefont {Franchini}}, \ and\ \bibinfo {author} {\bibfnamefont
  {G.}~\bibnamefont {Kresse}},\ }\href@noop {} {\bibfield  {journal} {\bibinfo
  {journal} {Scientific reports}\ }\textbf {\bibinfo {volume} {6}},\ \bibinfo
  {pages} {1} (\bibinfo {year} {2016})}\BibitemShut {NoStop}%
\bibitem [{\citenamefont {Deslippe}\ \emph
  {et~al.}(2012{\natexlab{b}})\citenamefont {Deslippe}, \citenamefont
  {Samsonidze}, \citenamefont {Strubbe}, \citenamefont {Jain}, \citenamefont
  {Cohen},\ and\ \citenamefont {Louie}}]{berkeleygw}%
  \BibitemOpen
  \bibfield  {author} {\bibinfo {author} {\bibfnamefont {J.}~\bibnamefont
  {Deslippe}}, \bibinfo {author} {\bibfnamefont {G.}~\bibnamefont
  {Samsonidze}}, \bibinfo {author} {\bibfnamefont {D.~A.}\ \bibnamefont
  {Strubbe}}, \bibinfo {author} {\bibfnamefont {M.}~\bibnamefont {Jain}},
  \bibinfo {author} {\bibfnamefont {M.~L.}\ \bibnamefont {Cohen}}, \ and\
  \bibinfo {author} {\bibfnamefont {S.~G.}\ \bibnamefont {Louie}},\ }\href
  {\doibase https://doi.org/10.1016/j.cpc.2011.12.006} {\bibfield  {journal}
  {\bibinfo  {journal} {Computer Physics Communications}\ }\textbf {\bibinfo
  {volume} {183}},\ \bibinfo {pages} {1269} (\bibinfo {year}
  {2012}{\natexlab{b}})}\BibitemShut {NoStop}%
\bibitem [{\citenamefont {Gulans}\ \emph {et~al.}(2014)\citenamefont {Gulans},
  \citenamefont {Kontur}, \citenamefont {Meisenbichler}, \citenamefont {Nabok},
  \citenamefont {Pavone}, \citenamefont {Rigamonti}, \citenamefont
  {Sagmeister}, \citenamefont {Werner},\ and\ \citenamefont
  {Draxl}}]{exciting}%
  \BibitemOpen
  \bibfield  {author} {\bibinfo {author} {\bibfnamefont {A.}~\bibnamefont
  {Gulans}}, \bibinfo {author} {\bibfnamefont {S.}~\bibnamefont {Kontur}},
  \bibinfo {author} {\bibfnamefont {C.}~\bibnamefont {Meisenbichler}}, \bibinfo
  {author} {\bibfnamefont {D.}~\bibnamefont {Nabok}}, \bibinfo {author}
  {\bibfnamefont {P.}~\bibnamefont {Pavone}}, \bibinfo {author} {\bibfnamefont
  {S.}~\bibnamefont {Rigamonti}}, \bibinfo {author} {\bibfnamefont
  {S.}~\bibnamefont {Sagmeister}}, \bibinfo {author} {\bibfnamefont
  {U.}~\bibnamefont {Werner}}, \ and\ \bibinfo {author} {\bibfnamefont
  {C.}~\bibnamefont {Draxl}},\ }\href {\doibase 10.1088/0953-8984/26/36/363202}
  {\bibfield  {journal} {\bibinfo  {journal} {Journal of Physics: Condensed
  Matter}\ }\textbf {\bibinfo {volume} {26}},\ \bibinfo {pages} {363202}
  (\bibinfo {year} {2014})}\BibitemShut {NoStop}%
\bibitem [{\citenamefont {Verdi}\ and\ \citenamefont
  {Giustino}(2015)}]{gius_verdi}%
  \BibitemOpen
  \bibfield  {author} {\bibinfo {author} {\bibfnamefont {C.}~\bibnamefont
  {Verdi}}\ and\ \bibinfo {author} {\bibfnamefont {F.}~\bibnamefont
  {Giustino}},\ }\href {\doibase 10.1103/PhysRevLett.115.176401} {\bibfield
  {journal} {\bibinfo  {journal} {Phys. Rev. Lett.}\ }\textbf {\bibinfo
  {volume} {115}},\ \bibinfo {pages} {176401} (\bibinfo {year}
  {2015})}\BibitemShut {NoStop}%
\bibitem [{\citenamefont {Baroni}\ \emph {et~al.}(2001)\citenamefont {Baroni},
  \citenamefont {de~Gironcoli}, \citenamefont {Dal~Corso},\ and\ \citenamefont
  {Giannozzi}}]{baroni_dfpt_2001}%
  \BibitemOpen
  \bibfield  {author} {\bibinfo {author} {\bibfnamefont {S.}~\bibnamefont
  {Baroni}}, \bibinfo {author} {\bibfnamefont {S.}~\bibnamefont
  {de~Gironcoli}}, \bibinfo {author} {\bibfnamefont {A.}~\bibnamefont
  {Dal~Corso}}, \ and\ \bibinfo {author} {\bibfnamefont {P.}~\bibnamefont
  {Giannozzi}},\ }\href {\doibase 10.1103/RevModPhys.73.515} {\bibfield
  {journal} {\bibinfo  {journal} {Rev. Mod. Phys.}\ }\textbf {\bibinfo {volume}
  {73}},\ \bibinfo {pages} {515} (\bibinfo {year} {2001})}\BibitemShut
  {NoStop}%
\bibitem [{\citenamefont {Giannozzi}\ and\ \citenamefont
  {Baroni}(2005)}]{giannozzi2005density}%
  \BibitemOpen
  \bibfield  {author} {\bibinfo {author} {\bibfnamefont {P.}~\bibnamefont
  {Giannozzi}}\ and\ \bibinfo {author} {\bibfnamefont {S.}~\bibnamefont
  {Baroni}},\ }in\ \href@noop {} {\emph {\bibinfo {booktitle} {Handbook of
  Materials Modeling: Methods}}}\ (\bibinfo  {publisher} {Springer},\ \bibinfo
  {year} {2005})\ pp.\ \bibinfo {pages} {195--214}\BibitemShut {NoStop}%
\bibitem [{\citenamefont {Kouba}\ \emph {et~al.}(2001)\citenamefont {Kouba},
  \citenamefont {Taga}, \citenamefont {Ambrosch-Draxl}, \citenamefont
  {Nordstr\"om},\ and\ \citenamefont {Johansson}}]{kouba2001}%
  \BibitemOpen
  \bibfield  {author} {\bibinfo {author} {\bibfnamefont {R.}~\bibnamefont
  {Kouba}}, \bibinfo {author} {\bibfnamefont {A.}~\bibnamefont {Taga}},
  \bibinfo {author} {\bibfnamefont {C.}~\bibnamefont {Ambrosch-Draxl}},
  \bibinfo {author} {\bibfnamefont {L.}~\bibnamefont {Nordstr\"om}}, \ and\
  \bibinfo {author} {\bibfnamefont {B.}~\bibnamefont {Johansson}},\ }\href
  {\doibase 10.1103/PhysRevB.64.184306} {\bibfield  {journal} {\bibinfo
  {journal} {Phys. Rev. B}\ }\textbf {\bibinfo {volume} {64}},\ \bibinfo
  {pages} {184306} (\bibinfo {year} {2001})}\BibitemShut {NoStop}%
\bibitem [{\citenamefont {Fr\"ohlich}(1954)}]{froehlich}%
  \BibitemOpen
  \bibfield  {author} {\bibinfo {author} {\bibfnamefont {H.}~\bibnamefont
  {Fr\"ohlich}},\ }\href {\doibase 10.1080/00018735400101213} {\bibfield
  {journal} {\bibinfo  {journal} {Advances in Physics}\ }\textbf {\bibinfo
  {volume} {3}},\ \bibinfo {pages} {325} (\bibinfo {year} {1954})}\BibitemShut
  {NoStop}%
\bibitem [{\citenamefont {Giustino}\ \emph {et~al.}(2007)\citenamefont
  {Giustino}, \citenamefont {Cohen},\ and\ \citenamefont
  {Louie}}]{giustino_eph_wannier_2007}%
  \BibitemOpen
  \bibfield  {author} {\bibinfo {author} {\bibfnamefont {F.}~\bibnamefont
  {Giustino}}, \bibinfo {author} {\bibfnamefont {M.~L.}\ \bibnamefont {Cohen}},
  \ and\ \bibinfo {author} {\bibfnamefont {S.~G.}\ \bibnamefont {Louie}},\
  }\href {\doibase 10.1103/PhysRevB.76.165108} {\bibfield  {journal} {\bibinfo
  {journal} {Phys. Rev. B}\ }\textbf {\bibinfo {volume} {76}},\ \bibinfo
  {pages} {165108} (\bibinfo {year} {2007})}\BibitemShut {NoStop}%
\bibitem [{\citenamefont {Hybertsen}\ and\ \citenamefont
  {Louie}(1986{\natexlab{b}})}]{hybrtsen_louie_1986}%
  \BibitemOpen
  \bibfield  {author} {\bibinfo {author} {\bibfnamefont {M.~S.}\ \bibnamefont
  {Hybertsen}}\ and\ \bibinfo {author} {\bibfnamefont {S.~G.}\ \bibnamefont
  {Louie}},\ }\href {\doibase 10.1103/PhysRevB.34.5390} {\bibfield  {journal}
  {\bibinfo  {journal} {Phys. Rev. B}\ }\textbf {\bibinfo {volume} {34}},\
  \bibinfo {pages} {5390} (\bibinfo {year} {1986}{\natexlab{b}})}\BibitemShut
  {NoStop}%
\bibitem [{\citenamefont {Puschnig}\ and\ \citenamefont
  {Ambrosch-Draxl}(2002)}]{puschnig_draxl_2002}%
  \BibitemOpen
  \bibfield  {author} {\bibinfo {author} {\bibfnamefont {P.}~\bibnamefont
  {Puschnig}}\ and\ \bibinfo {author} {\bibfnamefont {C.}~\bibnamefont
  {Ambrosch-Draxl}},\ }\href {\doibase 10.1103/PhysRevB.66.165105} {\bibfield
  {journal} {\bibinfo  {journal} {Phys. Rev. B}\ }\textbf {\bibinfo {volume}
  {66}},\ \bibinfo {pages} {165105} (\bibinfo {year} {2002})}\BibitemShut
  {NoStop}%
\bibitem [{\citenamefont {Perdew}\ \emph {et~al.}(2008)\citenamefont {Perdew},
  \citenamefont {Ruzsinszky}, \citenamefont {Csonka}, \citenamefont {Vydrov},
  \citenamefont {Scuseria}, \citenamefont {Constantin}, \citenamefont {Zhou},\
  and\ \citenamefont {Burke}}]{pbe_sol}%
  \BibitemOpen
  \bibfield  {author} {\bibinfo {author} {\bibfnamefont {J.~P.}\ \bibnamefont
  {Perdew}}, \bibinfo {author} {\bibfnamefont {A.}~\bibnamefont {Ruzsinszky}},
  \bibinfo {author} {\bibfnamefont {G.~I.}\ \bibnamefont {Csonka}}, \bibinfo
  {author} {\bibfnamefont {O.~A.}\ \bibnamefont {Vydrov}}, \bibinfo {author}
  {\bibfnamefont {G.~E.}\ \bibnamefont {Scuseria}}, \bibinfo {author}
  {\bibfnamefont {L.~A.}\ \bibnamefont {Constantin}}, \bibinfo {author}
  {\bibfnamefont {X.}~\bibnamefont {Zhou}}, \ and\ \bibinfo {author}
  {\bibfnamefont {K.}~\bibnamefont {Burke}},\ }\href {\doibase
  10.1103/PhysRevLett.100.136406} {\bibfield  {journal} {\bibinfo  {journal}
  {Phys. Rev. Lett.}\ }\textbf {\bibinfo {volume} {100}},\ \bibinfo {pages}
  {136406} (\bibinfo {year} {2008})}\BibitemShut {NoStop}%
\bibitem [{\citenamefont {Vorwerk}\ \emph {et~al.}(2019)\citenamefont
  {Vorwerk}, \citenamefont {Aurich}, \citenamefont {Cocchi},\ and\
  \citenamefont {Draxl}}]{Vorwerk_2019}%
  \BibitemOpen
  \bibfield  {author} {\bibinfo {author} {\bibfnamefont {C.}~\bibnamefont
  {Vorwerk}}, \bibinfo {author} {\bibfnamefont {B.}~\bibnamefont {Aurich}},
  \bibinfo {author} {\bibfnamefont {C.}~\bibnamefont {Cocchi}}, \ and\ \bibinfo
  {author} {\bibfnamefont {C.}~\bibnamefont {Draxl}},\ }\href {\doibase
  10.1088/2516-1075/ab3123} {\bibfield  {journal} {\bibinfo  {journal}
  {Electronic Structure}\ }\textbf {\bibinfo {volume} {1}},\ \bibinfo {pages}
  {037001} (\bibinfo {year} {2019})}\BibitemShut {NoStop}%
\bibitem [{\citenamefont {Alvertis}\ \emph
  {et~al.}(2023{\natexlab{b}})\citenamefont {Alvertis}, \citenamefont
  {Champagne}, \citenamefont {Del~Ben}, \citenamefont {da~Jornada},
  \citenamefont {Qiu}, \citenamefont {Filip},\ and\ \citenamefont
  {Neaton}}]{alvertis_nonuniform_bz_23}%
  \BibitemOpen
  \bibfield  {author} {\bibinfo {author} {\bibfnamefont {A.~M.}\ \bibnamefont
  {Alvertis}}, \bibinfo {author} {\bibfnamefont {A.}~\bibnamefont {Champagne}},
  \bibinfo {author} {\bibfnamefont {M.}~\bibnamefont {Del~Ben}}, \bibinfo
  {author} {\bibfnamefont {F.~H.}\ \bibnamefont {da~Jornada}}, \bibinfo
  {author} {\bibfnamefont {D.~Y.}\ \bibnamefont {Qiu}}, \bibinfo {author}
  {\bibfnamefont {M.~R.}\ \bibnamefont {Filip}}, \ and\ \bibinfo {author}
  {\bibfnamefont {J.~B.}\ \bibnamefont {Neaton}},\ }\href {\doibase
  10.1103/PhysRevB.108.235117} {\bibfield  {journal} {\bibinfo  {journal}
  {Phys. Rev. B}\ }\textbf {\bibinfo {volume} {108}},\ \bibinfo {pages}
  {235117} (\bibinfo {year} {2023}{\natexlab{b}})}\BibitemShut {NoStop}%
\bibitem [{\citenamefont {Dai}\ \emph {et~al.}(2024{\natexlab{b}})\citenamefont
  {Dai}, \citenamefont {Lian}, \citenamefont {Lafuente-Bartolome},\ and\
  \citenamefont {Giustino}}]{giustino_exc_polarons_24}%
  \BibitemOpen
  \bibfield  {author} {\bibinfo {author} {\bibfnamefont {Z.}~\bibnamefont
  {Dai}}, \bibinfo {author} {\bibfnamefont {C.}~\bibnamefont {Lian}}, \bibinfo
  {author} {\bibfnamefont {J.}~\bibnamefont {Lafuente-Bartolome}}, \ and\
  \bibinfo {author} {\bibfnamefont {F.}~\bibnamefont {Giustino}},\ }\href
  {\doibase 10.1103/PhysRevB.109.045202} {\bibfield  {journal} {\bibinfo
  {journal} {Phys. Rev. B}\ }\textbf {\bibinfo {volume} {109}},\ \bibinfo
  {pages} {045202} (\bibinfo {year} {2024}{\natexlab{b}})}\BibitemShut
  {NoStop}%
\end{thebibliography}%
\vspace{10cm}
 
\end{document}